\documentclass[letter]{article}
\usepackage{jheppub}
\usepackage{amsmath,amsfonts,relsize}
\usepackage[nameinlink,capitalize]{cleveref} 

\usepackage{graphicx}
\usepackage{dcolumn}
\usepackage[mathlines]{lineno}
\usepackage[caption=false]{subfig}
\usepackage{diagbox}
\usepackage{tabularx}
\usepackage{bm,color,xcolor}

\usepackage{floatrow}
\usepackage{comment}

\usepackage[rightcaption]{sidecap}
\sidecaptionvpos{figure}{c}
 
\newfloatcommand{capbtabbox}{table}[][\FBwidth]

\usepackage{enumitem}

\usepackage{xspace}

\usepackage{slashed,physics} 
\usepackage{float}
\usepackage{multirow}
\usepackage{mathtools}
\usepackage{appendix}
\usepackage[colorlinks=true
,urlcolor=blue
,anchorcolor=blue
,citecolor=blue 
,filecolor=blue
,linkcolor=blue
,menucolor=blue
,linktocpage=true
,pdfproducer=medialab
,pdfa=true
]{hyperref}
\usepackage{bm,color,color,soul}
\definecolor{UMNRed}{RGB}{144, 0, 33}
\definecolor{UMNGold}{RGB}{255 204 51}
\definecolor{CaltechOrange}{RGB}{255  108 12}
\usepackage{lineno}

\newcommand{\omegaTrans}{\omega}

\newcommand{\beq}{\begin{equation}}
\newcommand{\eeq}{\end{equation}}
\newcommand{\bea}{\begin{eqnarray}}
\newcommand{\eea}{\end{eqnarray}}

\title{Probing Millicharged Particles at an Electron Beam Dump with Ultralow-Threshold Sensors
}

\author[a]{Rouven Essig,}
\author[b]{Peiran Li,}
\author[b]{Zhen Liu,}
\author[a,c]{Megan McDuffie,}
\author[d]{Ryan Plestid,}
\author[a,c]{Hailin Xu}

\affiliation[a]{C.N. Yang Institute for Theoretical Physics, Stony Brook University, Stony Brook, NY 11794}
\affiliation[b]{School of Physics and Astronomy, University of Minnesota, Minneapolis, MN 55455, USA}
\affiliation[c]{Physics and Astronomy Department, Stony Brook University, NY 11794, USA}
\affiliation[d]{Walter Burke Institute for Theoretical Physics, \\ California Institute of Technology, Pasadena, CA 91125, USA}

\emailAdd{rouven.essig@stonybrook.edu}
\emailAdd{li001800@umn.edu}
\emailAdd{zliuphys@umn.edu}
\emailAdd{megan.mcduffie@stonybrook.edu}
\emailAdd{rplestid@caltech.edu}
\emailAdd{hailin.xu@stonybrook.edu}

\preprint{\small CALT-TH/2024-048,~YITP-SB-2024-30,~UMN-TH-4407/24}

\abstract{
We propose to search for millicharged particles produced in high-intensity electron beam dumps using small ultralow-threshold sensors. 
As a concrete example, we consider a Skipper-CCD placed behind the beam dump in Hall~A at Jefferson Lab.  We compute the millicharged particle flux, including both electromagnetic cascade and meson productions emanating from an aluminum target. We find that the sensitivity of a modest $2\times 14$ array of Skipper-CCDs can exceed the sensitivity of all existing searches for millicharged particle masses below 1.5~GeV, and is either competitive or world leading when compared to other proposed experiments. Our results demonstrate that small-scale ultralow threshold silicon devices can enhance the reach of accelerator-based experiments, while fitting comfortably within existing experimental halls. 
}

\begin{document}
\maketitle

\section{Introduction}

Extensions of the Standard Model (SM) are well 
motivated due, for example, to the observed dark matter relic abundance of our universe, the presence of neutrino masses, and the observed baryon asymmetry of the universe. New physics may be strongly coupled, in which case it must be heavy and is best searched for at high-energy colliders.  However, extremely feebly coupled dark sectors are viable even for masses smaller than $\sim 1~{\rm GeV}$. In this regime, beyond the SM (BSM) physics can be efficiently searched for in high-intensity accelerator experiments~\cite{Bjorken:2009mm,Prinz:1998ua,LSND:2001akn,Magill:2018tbb,Kahn:2014sra,Kim:2021eix,Arguelles:2019xgp,SENSEI:2023gie}. 

A simple extension of the Standard Model is a particle with a small electric charge (e.g., a particle with small hypercharge that is a singlet under $SU(2)$). Such particles are often referred to as millicharged particles (mCPs); they have interesting cosmological signatures and, if they compose the dark matter, novel direct-detection prospects~\cite{Essig:2011nj,Essig:2015cda,Crisler:2018gci,SENSEI:2019ibb,SENSEI:2020dpa,DarkSide:2018ppu,DarkSide:2022knj,FUNKExperiment:2020ofv,DAMIC:2019dcn,XENON:2019gfn,Essig:2017kqs,EDELWEISS:2020fxc,SuperCDMS:2018mne}. The existence of mCPs has interesting consequences related to charge quantization and grand unified theories~\cite{Okun:1983vw,Brahm:1989jh}, and ``effective mCPs'' naturally arise via kinetic mixing in the presence of a light, but massive, dark photon~\cite{Holdom:1985ag,Galison:1983pa}.

Searches for mCPs using accelerator facilities \cite{Prinz:1998ua,Balasubramanian:2023pap,LSND:2001akn,Magill:2018tbb,Harnik:2019zee,Kahn:2014sra,Arguelles:2019xgp,ArgoNeuT:2019ckq,Kim:2021eix,CMS:2012xi,Haas:2014dda,Foroughi-Abari:2020qar,milliQan:2021lne,SENSEI:2023gie} or cosmic-ray production \cite{Plestid:2020kdm,MACRO:2000bht,Majorana:2018gib,ArguellesDelgado:2021lek,Du:2022hms,Wu:2024iqm} are completely agnostic to astrophysical assumptions that can hamper searches for mCP dark matter. They can, therefore, easily probe models with an mCP in their spectrum without any astrophysical, cosmological, or model-dependent uncertainties. Currently, the strongest constraints below $\sim 2 ~{\rm GeV}$ in mass come from a combination of accelerator-based experiments~\cite{SENSEI:2023gie,Marocco:2020dqu}. 
The strongest published constraints on mCPs with masses below $\sim 100~{\rm MeV}$ are obtained from the SLAC-mQ collaboration \cite{Prinz:1998ua}, a reanalysis of LSND's $\nu e \rightarrow \nu e$ scattering dataset~\cite{LSND:2001akn,Magill:2018tbb}, and a recent search by the SENSEI collaboration using data taken in the MINOS hall at Fermilab~\cite{SENSEI:2023gie}.   

The SLAC-mQ experiment used a 29.5~GeV electron beam on a 6~radiation-length target consisting of 75\% tantalum and 25\% rhenium, with a detector placed downstream at a distance of $82.6~{\rm m}$~\cite{Prinz:1998ua}. 
The success of the SLAC-mQ experiment strongly motivates a renewed study of high-intensity electron beam dumps. In particular, we are motivated by the proposed Beam Dump Experiment (BDX) in the Continuous Electron Beam Accelerator Facility (CEBAF) at Jefferson Lab~\cite{BDX:2014pkr,BDX:2016akw,BDX:2017jub,BDX:2019afh}.  The BDX collaboration plans to deliver $10^{22}$ electrons on target (EOT) over six months, which is three orders of magnitude larger than the $8.4 \times 10^{18}$~EOT delivered in the SLAC-mQ experiment. 

Besides increasing the EOT, we advocate here for using ultralow-threshold sensors. This is motivated by the fact that mCPs that scatter in a detector medium prefer to transfer very little energy to it.  Moreover, the SENSEI collaboration performed an opportunistic search for mCPs using their Skipper-CCD detector, which happened to be located downstream of the NuMI beam in the MINOS cavern at Fermilab~\cite{SENSEI:2023gie}.  Despite the fact that only a single 2-gram Skipper-CCD was in operation and the SENSEI detector was $\sim$1~km away from the target, the low-energy threshold of the Skipper-CCD enhances the sensitivity to mCPs compared to a typical detector that has a much higher threshold, and world-leading limits were placed on mCPs.    
The Oscura collaboration has proposed using a 1-kg Skipper-CCD detector to search for mCPs using the NuMI beam as part of an engineering test on the way to developing a 10-kg dark matter detector~\cite{Oscura:2023qch}. 
Skipper-CCDs have also been used in mCP searches produced in reactors~\cite{CONNIE:2024off}. 

In this paper, we study the sensitivity of small, low-threshold sensors placed downstream from an electron beam-dump facility. Following the success of the SENSEI search~\cite{SENSEI:2023gie}, Skipper-CCDs are an obvious example, but other low-threshold sensors used in sub-GeV dark matter searches are also good candidates. This includes the SuperCDMS HVeV detectors, which consist of transition edge sensor (TES) attached to a silicon crystal~\cite{SuperCDMS:2018mne,SuperCDMS:2020ymb,SuperCDMS:2024yiv}, or the detectors used by the EDELWEISS collaboration, such as a neutron-transmutation-doped (NTD) thermal sensor glued to a germanium crystal~\cite{EDELWEISS:2020fxc}.  For concreteness, we focus in this paper on a silicon detector.

A significant advantage of low-threshold detectors is that they are small and versatile, and relatively easy to deploy.  
We study different production modes, first matching and then exceeding the sophistication of the SLAC-mQ target simulation (i.e., we include new production modes). Next, we argue that a zero-background search with a small Skipper-CCD detector is possible and that the sensitivity of a BDX-like setup would have world-leading sensitivity to  mCPs across a wide range of masses. 

The rest of the paper is organized as follows. In \cref{sec:production}, we outline the dominant mCP production modes both in the first radiation length and in the ensuing electromagnetic cascade, including also production from meson decays.
Next, in \cref{sec:detection}, we discuss the detector scattering cross section for both high- and low-energy threshold detectors, emphasizing the role of the plasmon resonance for the latter. This leads naturally to \cref{sec:results}, where we present our results for the reach of a Skipper-CCD detector downstream of a BDX-like electron beam dump. In \cref{sec:conclusions}, we summarize our findings and outline potentially interesting future directions. Three appendices provide additional details. In \cref{app:energy-thesholds}, we discuss additional details for the mCP production. In~\cref{app:MCS}, we estimate the scattering angle of mCP interacting in the shield before reaching the detector volume. In~\cref{app:backgrounds}, we provide estimates of beam-and cosmic-ray induced backgrounds.

\section{Millicharged-particle production in electron beam dumps \label{sec:production} }

In this section, we calculate the flux of mCPs emanating from an electron beam-dump target before discussing the detection of mCPs in \cref{sec:detection}. For concreteness, we use benchmarks that are motivated by BDX: $10^{22}$~EOT, a 10.6~GeV electron-beam energy, and a 300~cm thick aluminum target~\cite{BDX:2014pkr,BDX:2016akw}.  
\begin{figure}[t!]
    \centering
    \includegraphics[width=0.34\textwidth]{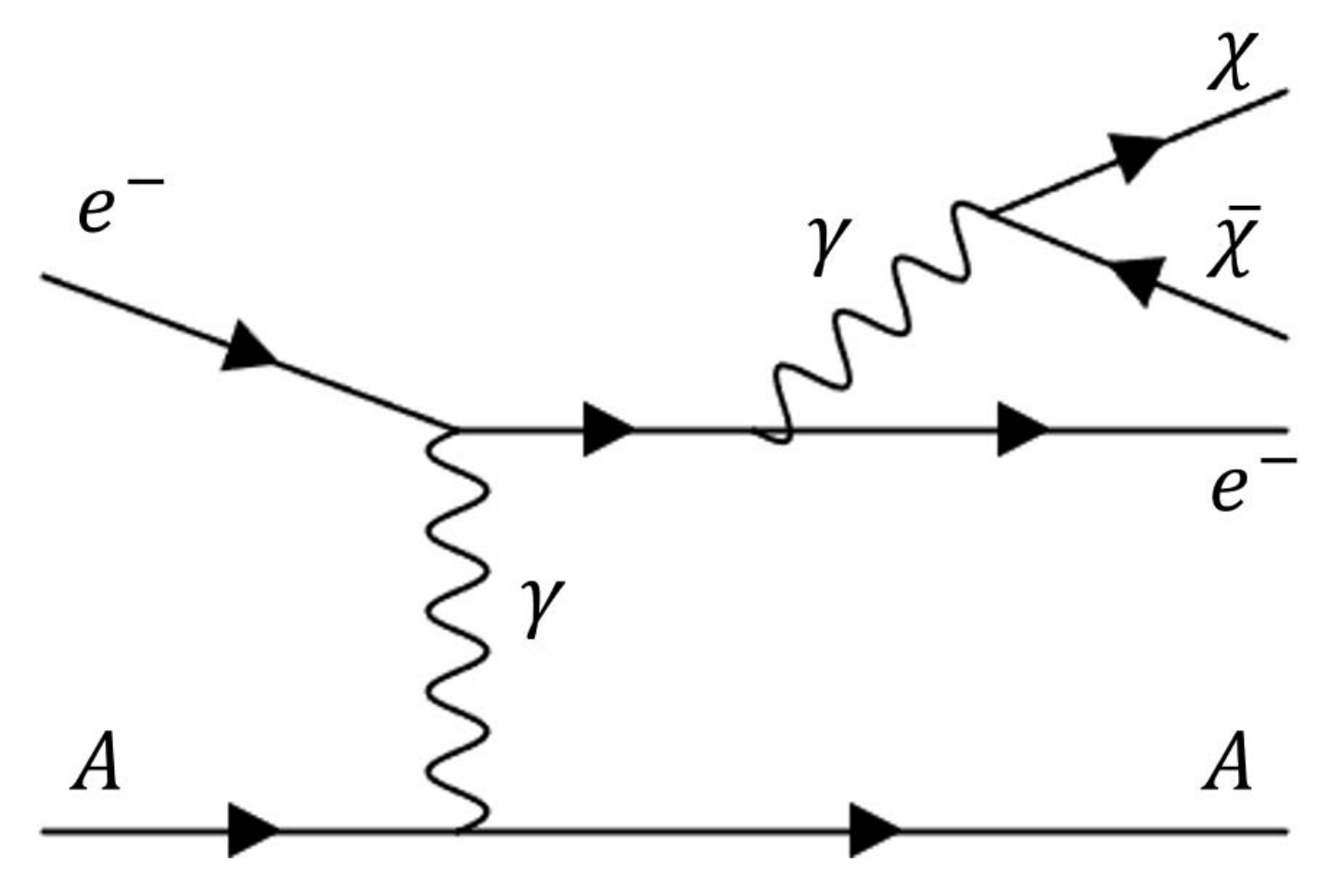}~
    ~~~~~~\includegraphics[width=0.34\textwidth]{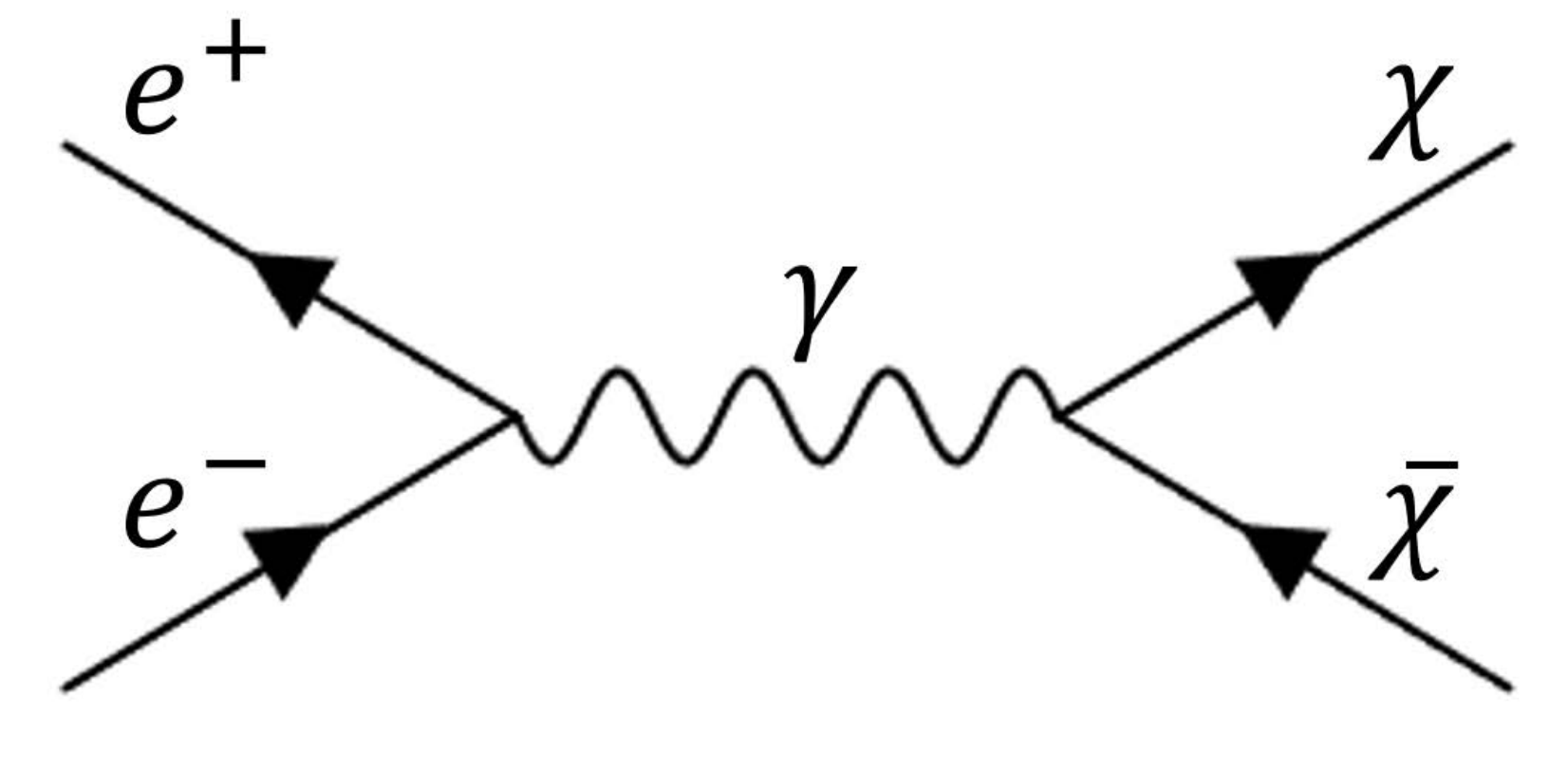}\\\vskip 3mm
    \includegraphics[width=0.34\textwidth]{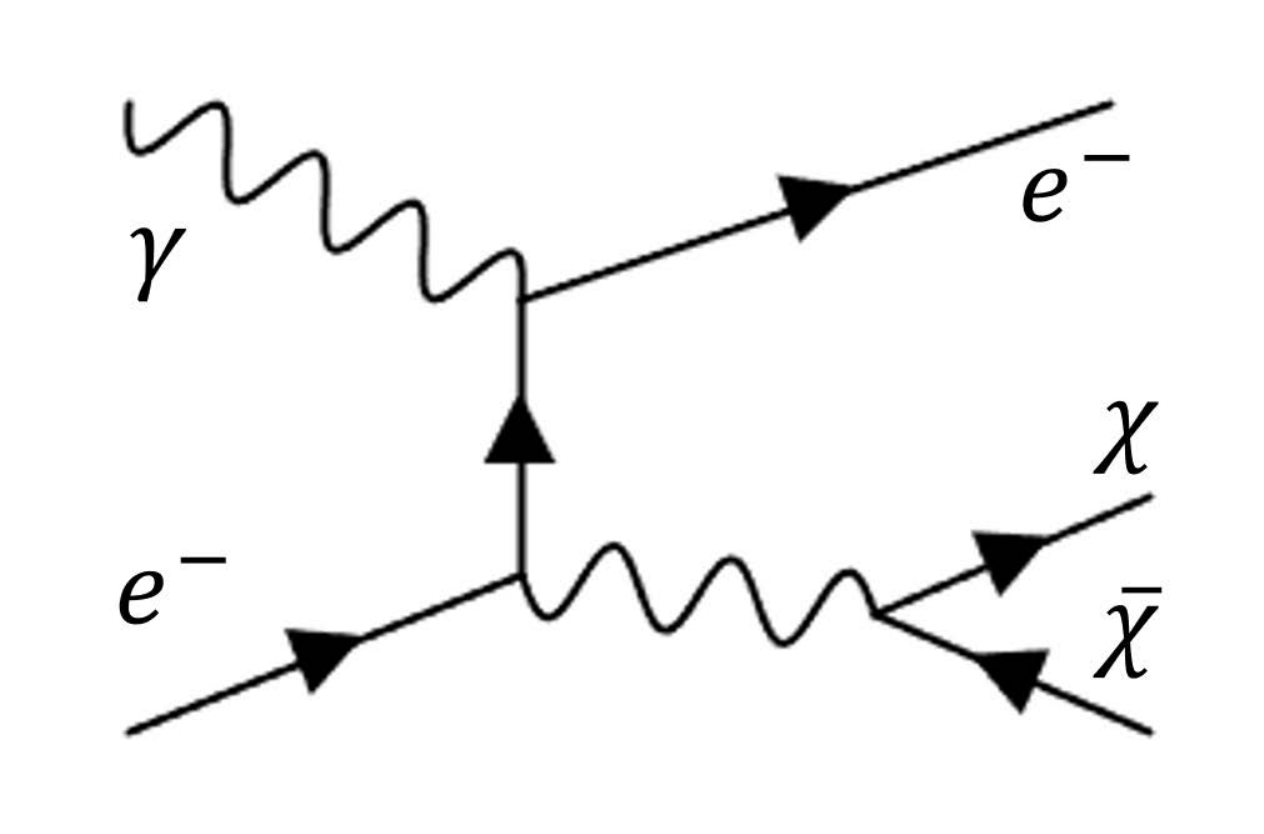}~
    ~~~~~~\includegraphics[width=0.34\textwidth]{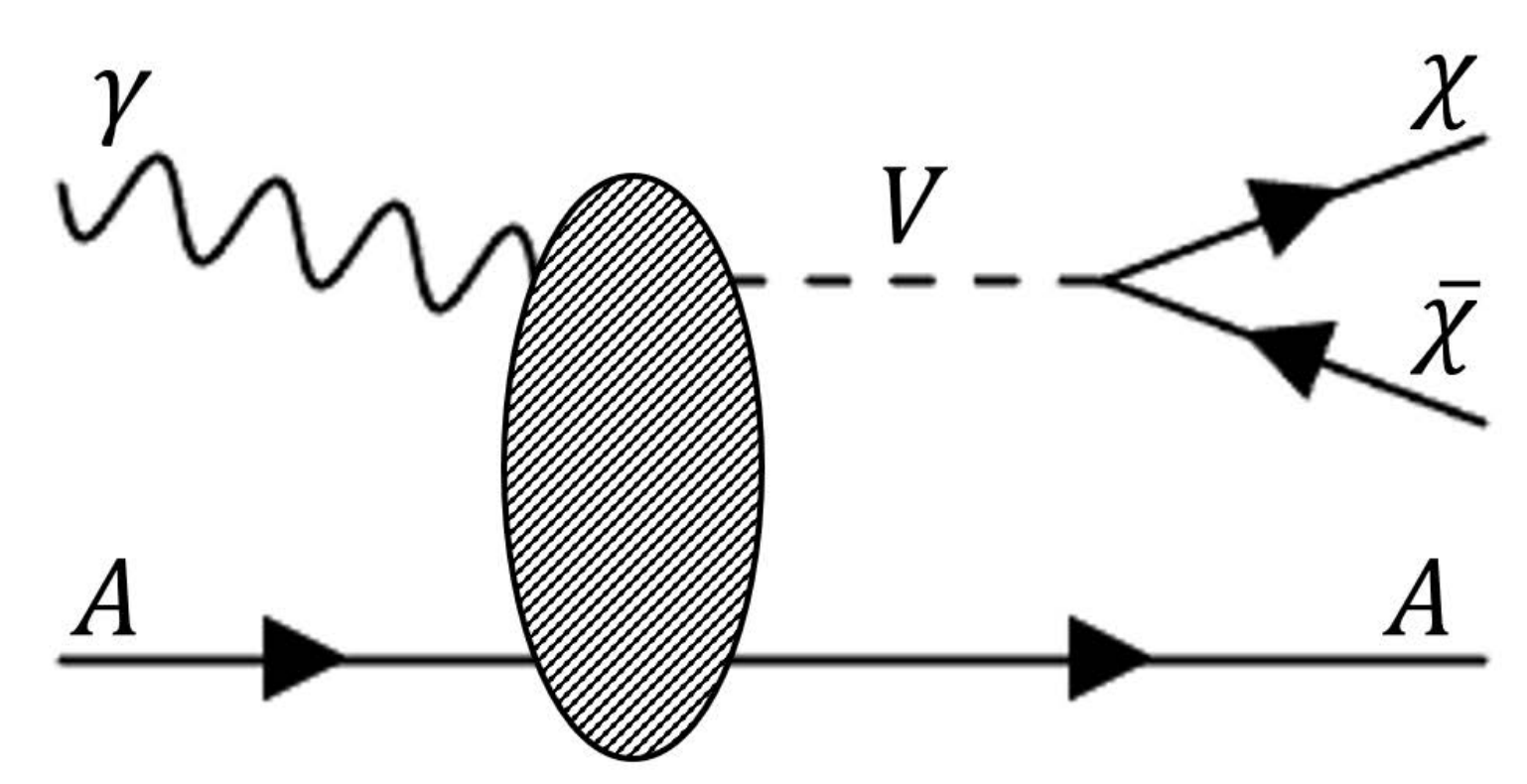}
    \caption{The leading Feynman diagrams for the production of mCPs ($\chi$) in an electron beam incident on a target containing nuclei $A$.  \textbf{Top left:} trident production ($e^-A\to e^-A\chi\bar{\chi}$); \textbf{Top right:} electron-positron annihilation ($e^+ e^-\to\chi\bar{\chi}$);  
    \textbf{Bottom left:} Compton production ($\gamma e^-\to e^- \chi\bar{\chi}$);  
    \textbf{Bottom right:} vector meson production with decay ($\gamma A\to A \chi\bar{\chi}$), which we treat in the narrow-width approximation. Both electron-positron annihilation production and Compton production involve secondary positrons and photons from the electromagnetic shower.
    \label{fig:feynman_diagram_production}}
\end{figure}

We organize the calculation as follows: First, we discuss trident production of mCPs, $\chi$ and $\bar{\chi}$, in an electron ($e^-$) beam interacting with aluminum nuclei $A$ ($e^- A \rightarrow e^- A \chi\bar{\chi}$) in the first radiation length of the target. Next, we consider the electromagnetic cascade and production stemming from daughter photons, electrons, and positrons. This matches the sophistication of the SLAC-mQ experiment for the trident production channel , and further includes two new production mechanisms:
positron annihilation ($e^+e^- \rightarrow \chi\bar{\chi}$) and Compton production ($\gamma e \rightarrow e \chi\bar{\chi}$). We find that the production of mCP via annihilation can dominate over the trident mechanism in certain regions of parameter space (for $2~{\rm MeV}\lesssim m_\chi \lesssim 50~{\rm MeV}$); however, we also find that the Compton production is always subdominant. Finally, we estimate the photonuclear production channels, which dominates the flux at larger mCP masses (specifically for $m_\chi \gtrsim 130~{\rm MeV}$). We show the leading Feynman diagrams for the dominant four production modes in \cref{fig:feynman_diagram_production}. 

\subsection{Production of mCPs in the first radiation length}
In the first radiation length, the leading channel of mCP production from the 10.6~GeV electron beam is the trident process
\[
    e^-A\to e^-A\chi\bar{\chi} \,,
\]
where $A$ is the aluminum nucleus. One of the leading Feynman diagram topologies is shown in~\cref{fig:feynman_diagram_production} (top-left). The computation of this $2\rightarrow4$ process is performed using {\tt MadGraph5\_aMC@NLO}~\cite{Alwall:2014hca}. The electromagnetic coupling between the Al nucleus and mCP is implemented via {\tt FeynRules}~\cite{Alloul:2013bka} with an atomic form factor (the terms involving $a$ and $a^\prime$), a nuclear form factor (the term involving $d$) for coherent scattering, and a nucleon form factor (the term involving $\mu_p$) for incoherent scattering~\cite{Bjorken:2009mm,Celentano:2020vtu,Kim:1973he,Tsai:1973py}, as follows
\begin{align*}
    |F(t)|^2=Z^2\left(\frac{a^2 t}{1+a^2 t}\right)^2\left(\frac{1}{1+t/d}\right)^2+Z\left(\frac{a'^2 t}{1+a'^2 t}\right)^2\left(\frac{1+\frac{t}{4m_p^2}(\mu_p^2-1)}{(1+\frac{t}{0.71~\text{GeV}^2})^4}\right)\,.
\end{align*}
Here, $Z$ is the atomic number, $a=111~Z^{-1/3}/m_e$, $a'=773~Z^{-2/3}/m_e$, $d=0.164~\text{GeV}^2~A^{-2/3}$ with atomic mass number $A$, $\mu_p=2.79$, $m_p=0.938~{\rm GeV}$ is the proton mass, and $t=-q^2$ with $q^2$ the four-momentum transfer squared (note that $t\geq 0$).

As a useful benchmark against other lepton beam-dump studies in the literature~\cite{Diamond:2013oda,Chu:2018qrm}, we consider the mCP flux produced in the first radiation length. This approach offers a conservative lower-bound on the mCP flux and is a useful approximation at the order-of-magnitude level. As we will see below, including the full electromagnetic cascade and other production channels further increases the flux of mCPs. 

\begin{figure}[t!]
    \centering
    \includegraphics[width=0.49\textwidth]{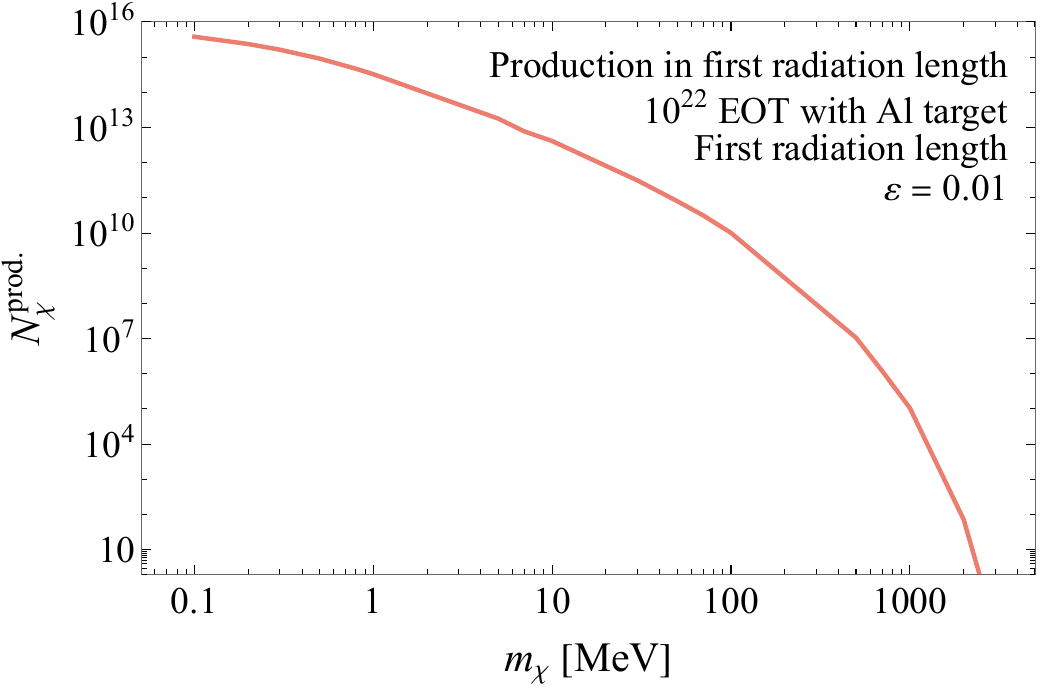}~
    \includegraphics[width=0.49\textwidth]{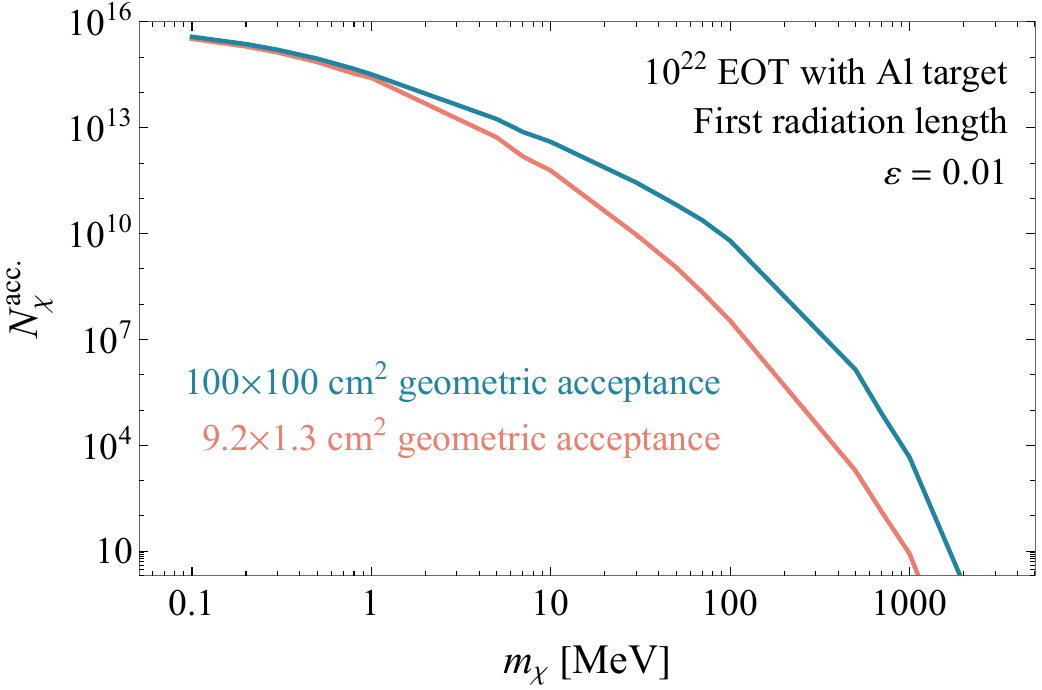}
    \caption{\textbf{Left:} The number of mCPs produced  from trident production $e^-N\to e^-N\chi\bar{\chi}$ in the first radiation length  versus the mCP mass $m_\chi$ from a 10.6~GeV electron beam incident on a thick aluminum target. The millicharge is chosen to be  $\varepsilon=0.01$ (in units of the electron charge). \textbf{Right:} The number of mCPs produced from trident production in the first radiation length and passing through a $9.2 \times 1.3~\text{cm}^2$ or $100 \times 100~\text{cm}^2$ area along the beam center 20~m downstream of the target, where our representative silicon detector will be located.
    \label{fig:first_radiation_number}}
\end{figure}

The total number of mCPs produced in the first radiation length is computed using 
\begin{align}
    N_\chi^\text{prod.}=2\cdot N_\text{EOT}\cdot \frac{\rho \cdot N_A}{M_\text{atom}}\cdot X_0\cdot\sigma_\text{prod.}\,,
\end{align}
where $N_\text{EOT}=10^{22}$ is the number of electrons on target~\cite{BDX:2019afh}, $X_0=8.89~\text{cm}$ is the radiation length of aluminum~\cite{ParticleDataGroup:2024cfk}, 
$\rho=2.7~\text{g}/\text{cm}^3$ is the density of the aluminum target, $M_\text{atom}=27~\text{g}/\text{mol}$ is the molar mass of aluminum, $N_A$ is the Avogadro constant, and $\sigma_\text{prod.}$ is the cross section of the trident production with a $10.6$~GeV electron beam. 

In the first-radiation-length approximation, the total number of mCPs and the number of mCPs within the detector geometric acceptance are shown in the left and right plot of \cref{fig:first_radiation_number}, respectively. The geometric acceptance assumes a detector placed 20~m downstream on the beam line with rectangular areas of $9.2~{\rm cm}\times 1.3~{\rm cm}$ and $1~{\rm m} \times 1~{\rm m}$; these correspond, respectively, to the size of a single Skipper-CCD used in the SENSEI search~\cite{SENSEI:2023gie}, and the transverse dimensions of the nominal BDX detector \cite{BDX:2019afh}.  As $m_\chi$ increases, the signal cross section decreases. Furthermore, heavier mCPs are less forward through trident production and, hence, have a reduced geometric acceptance.

\subsection{Production of mCPs in the electromagnetic cascade}
The electromagnetic (EM) cascade includes all of the secondary Quantum Electrodynamics (QED) particles that are generated when the incident 10.6~GeV electron passes through the target. This includes photons, which are emitted as bremsstrahlung, and all of their subsequent descendants, which are electrons, positrons, and photons. The mCPs generated by these three particles can be classified into three production modes---trident production, annihilation of positrons with atomic electrons, and Compton scattering. 
We discuss each of them below, but note that Compton scattering is subdominant; the leading Feynman diagram for each process is shown in \cref{fig:feynman_diagram_production}.
We simulate the SM EM cascade using {\tt PETITE}~\cite{Blinov:2024pza}, combined with a set of {\tt MadGraph} simulations for the mCP production channels.

\subsubsection*{Trident production in a nuclear field}

A single 10.6~GeV electron on target generates from the EM shower about 200~photons, 450~electrons, and 300~positrons, each with energy above 10~MeV.\footnote{Note that this counting is sensitive to the infrared cut off, and we count every particle once per (forward-dominant) scattering in the Monte Carlo.} 
Just like the primary electron beam, the secondary electrons and positrons can produce mCPs from trident processes. For each electron/positron in the shower, the number of mCPs from trident production is
\begin{align}
    N=2\cdot\frac{\rho \cdot N_A}{M_\text{atom}}\cdot \lambda_{\rm MFP} \cdot\sigma(E_e)\,,
\end{align}
where $\lambda_{\rm MFP}(E_e)$ is the mean-free path of the electron/positron, and $\sigma(E_e)$ is the production cross section, both for an incident electron/positron with energy $E_e$. The cross section $\sigma(E_e)$ is computed via {\tt Madgraph}. 
For large mCP masses, as the energy decreases, the cross section decreases due to phase space suppression. For small mCP masses, the cross section also drops as the energy decreases, which can be understood within the context of the Weizs\"acker-Williams approximation ~\cite{Bjorken:2009mm,Liu:2017htz,Gninenko:2018ter}.

\begin{figure}[t!]
    \centering
    \includegraphics[width=0.49\textwidth]{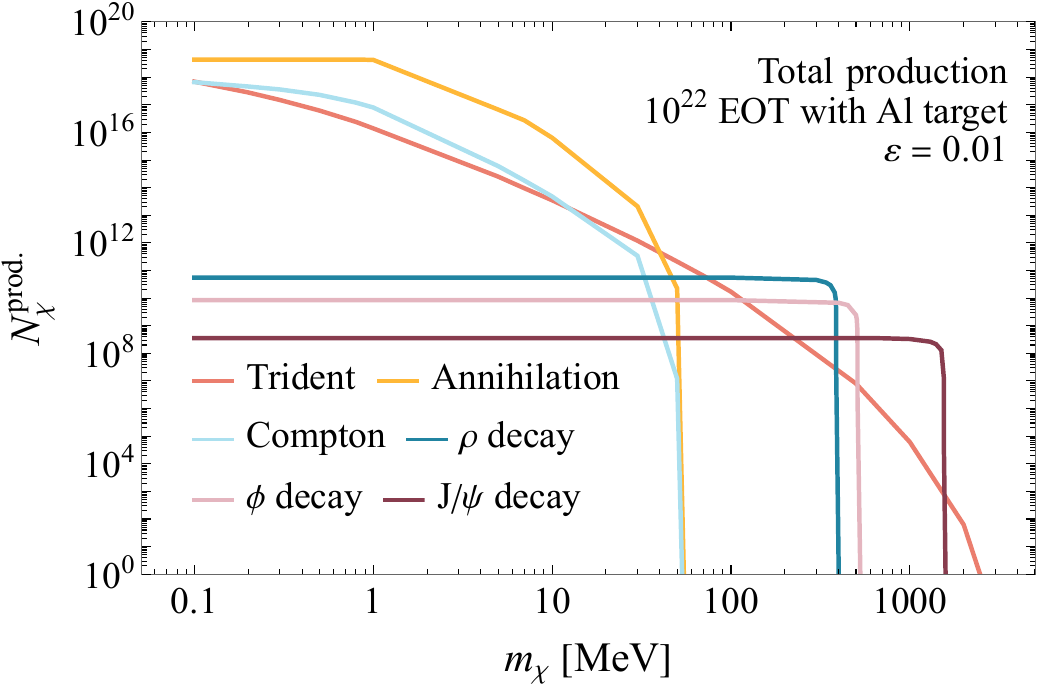}\\
    \includegraphics[width=0.49\textwidth]{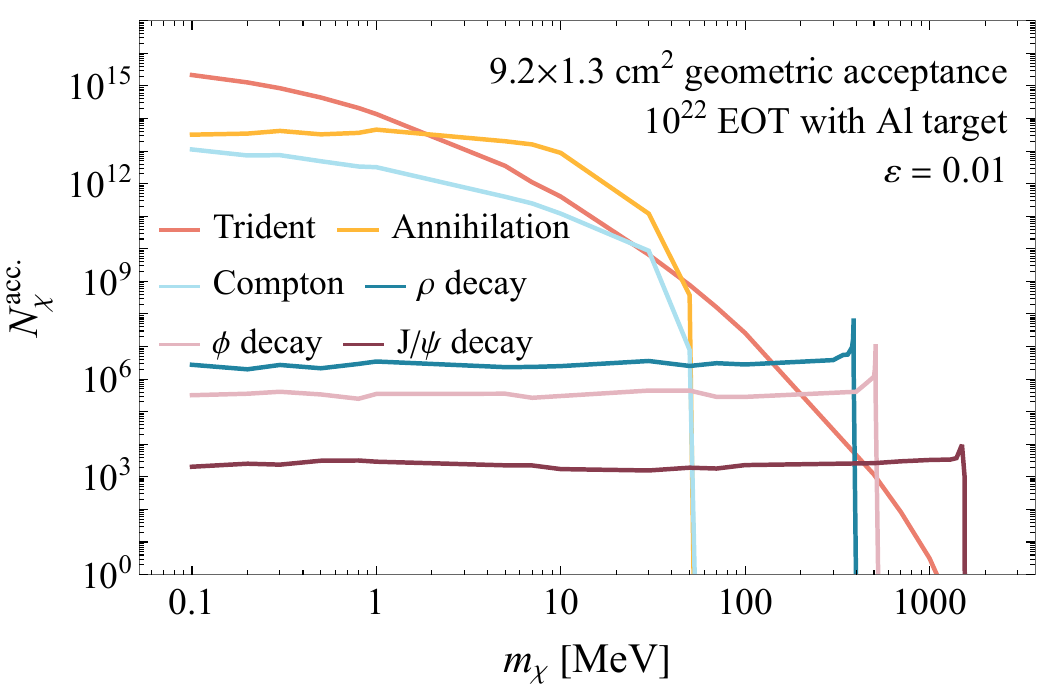}~
    \includegraphics[width=0.49\textwidth]{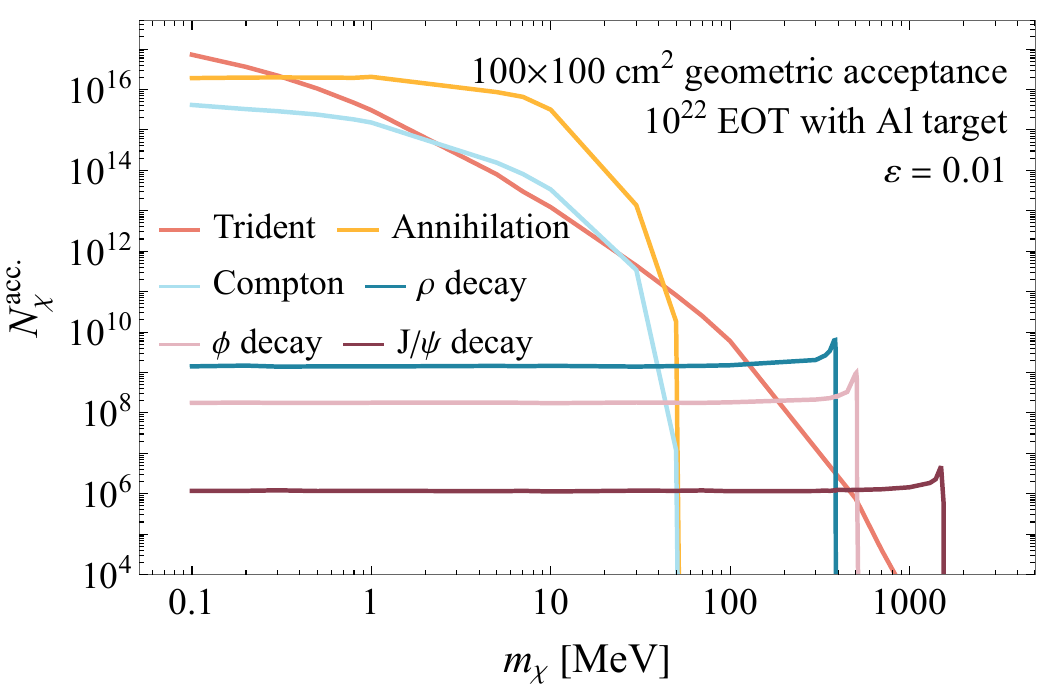}
    \caption{ \textbf{Top:} The number of mCPs produced versus the mCP mass $m_\chi$ from a 10.6~GeV electron beam incident on a thick aluminum target.
    The millicharge is chosen to be  $\varepsilon=0.01$ (in units of the electron charge). The mCPs are produced in trident processes (red), electron-positron annihilation (yellow), Compton scattering (light blue), and various meson decays ($\rho$ (blue), $\phi$ (pink), and $J/\Psi$ (dark red)).  The inclusion of the EM cascade increases the trident yield by a few (roughly $\sim 2\times$ more mCPs) compared to the mCPs produced in the first radation length (see \cref{fig:first_radiation_number}), whereas $e^+e^-$ annihilation and meson decays introduce qualitatively new features that dominate the flux for different $m_\chi$. 
    \textbf{Bottom left:} The total number of mCPs that pass through an area of $9.2 \times 1.3~ \text{cm}^2$ at a distance 20~m downstream of the target on the beam line, where our representative silicon detector will be located.  The sharp ``peaks'' near the meson thresholds arise from the increased acceptance due to the small $p_T$ imparted in the meson decay to the mCP pairs. 
    \textbf{Bottom right:} The total number of mCPs that pass through an area of $100 \times 100~\text{cm}^2$ at a distance of 20~m downstream of the target on the beam line. 
    }
    \label{fig:shower_production_number}
\end{figure}

In general, as the energy of the secondary particles decreases, the mCPs are produced with larger angles relative to the beam line. Therefore, mCPs that originate from low-energy secondary particles will have a small geometric acceptance at the detector. In our analysis, we ignore all secondary particles that have an energy below 10~MeV. We checked that the mCPs produced from secondary particles with energies below 10~MeV are negligible, and we provide the relevant details in \cref{app:energy-thesholds}. The number of mCPs produced in trident processes for different mCP masses $m_\chi$ are shown with the red curves in \cref{fig:shower_production_number}, both without (left plot) and with (right plot) the detector acceptance.   

Given that we can ignore secondary particles with energy below 10~MeV, we note that the precise thickness of the aluminum target (which is 300~cm) is not important, as long as it is thicker than about 10~radiation lengths ($\sim$90~cm), as after 10 radiation lengths, the secondary particles will have an average energy of $\sim 10~{\rm GeV} \times (1/2)^{10} \lesssim 10~{\rm MeV}$.

\subsubsection*{Annihilation of positrons with atomic electrons}
Inside a material, high-energy photons pair-produce electrons and positrons in the Coulomb field of nuclei. When positrons propagate through the material, they can annihilation to mCPs, $e^+ e^-\to \chi\bar{\chi}$.  This is an important production mechanism, since the cross section per atom is proportional to $Z \times (\alpha^2 \varepsilon^2)$ as compared to $\alpha^4 Z^2$ for trident production, where $\alpha\simeq 1/137$ is the fine-structure constant. The annihilation cross section is
\begin{align*}
    \sigma_\text{anni.}=\frac{4\pi\alpha^2 \varepsilon^2}{3}\frac{\left(s+2m_e^2\right)}{s^3\sqrt{s-4m_e^2}}\left(s+2m_\chi^2\right)\sqrt{s-4m_\chi^2}\,,
\end{align*}
where $s=2m_e E+m_e^2$ is the center-of-mass-energy squared. The production rate is proportional to $s^{-1}$ for light mCPs. For heavy mCPs, the cross section will rapidly decrease near the kinematic threshold. We show the mCP production from annihilation in the yellow line in \cref{fig:shower_production_number}, both without (left) and with (right) the detector acceptance, which exhibits these features clearly. Furthermore, looking at the right panel of \cref{fig:shower_production_number}, the annihilation process is a dominant channel for $m_\chi$ between 2~MeV to 50~MeV. This is because the annihilation process produces mCPs at smaller angles compared to the trident process, which are more spread out.
For a given incident positron with momentum $p_z$, the boost of the mCPs from electron-positron annihilation is the three-momentum of the photon propagator, which also equals $p_z$. As $m_\chi$ increases, the number of mCPs traversing the detector even increases slightly, since the transverse momentum of the mCP decreases.

\subsubsection*{Compton production off atomic electrons}
Photons produced in the EM cascade can Compton scatter off atomic electrons and produce mCPs.  We find that Compton production of mCPs is subdominant to the mCPs produced in both trident processes and electron-positron annihilation. The production cross section is computed numerically via {\tt MadGraph} in this study. To understand the result, here we provide a parametric estimation via the annihilation process convolved with a splitting function, 
\begin{align*}
    \sigma_\text{Comp.}=\frac{\alpha}{2\pi}\ln{\left(\frac{s}{m_e^2}\right)}\int_{x_\text{min}}^1 \text{d}x~P_{\gamma \to ee}(x)\sigma_\text{anni.}(xs)\,,
\end{align*}
where $P_{\gamma\to ee}(x)=x^2+(1-x)^2$ is the splitting function for the incident photon to produce an electron-positron pair, and $x_\text{min}\approx 4m_\chi^2/s$. Both annihilation and Compton production have the same energy threshold, which is set by $m_\chi$, as shown explicitly in \cref{fig:shower_production_number}. The leading-logarithm $\ln\left(\frac{s}{m_e^2}\right)$ provides a factor of $\mathcal{O}(10)$ enhancement, but $\frac{\alpha}{2\pi}$ gives a suppression of $\sim 10^{-3}$. Thus, the Compton production is generally smaller than annihilation by about two orders of magnitude.

\subsection{Production of mCPs in meson decays} 
\label{subsec:mesons}
As a final production mode, we include an estimate of photonuclear reactions that source mesons, which subsequently decay into $\chi\bar{\chi}$ pairs; we specifically focus on the production of vector mesons.
We have verified that the yield of pseudoscalar mesons produced via Primakoff scattering is subdominant to the electromagnetic cascade across the full range of available mCP masses. In contrast, vector-meson production is significantly larger than pseudoscalar meson production, which is in agreement with estimates produced for LDMX and NA64~\cite{Schuster:2021mlr}. 
Moreover, vector mesons can produce more mCPs than those produced in the EM cascade, at least for $m_\chi \gtrsim 130$~MeV. 
We, therefore, study $\gamma \rightarrow \rho, \phi , J/\psi$ production modes.  

We focus first on the photoproduction of $\rho$ mesons. Since the $\rho$ mass is relatively heavy, production occurs dominantly in the first radiation length of the target. The $\rho$ is produced from an approximately equal mixture of coherent and incoherent processes. For simplicity, in the discussion that follows, we focus on the coherent channel; however, we include both coherent and incoherent production (i.e., on individual nucleons) in our estimates. We only include photons from the first radiation length, which will underestimate the full flux by a factor of a few.

In the complete screening approximation (appropriate for $E_e\simeq 10~{\rm GeV}$), the number of photons produced per radiation length per energy interval $\dd E_\gamma$  is given by~\cite{Tsai:1966js} 
\begin{equation}
    \frac{\dd N_\gamma}{\dd E_\gamma} \simeq \frac{N_{\rm EOT}}{E_\gamma} \qty[ \frac43\qty(1-\tfrac{E_\gamma}{E_e})^2 + \qty(\tfrac{E_\gamma}{E_e})^2]\,. 
\end{equation}
We multiply this expression by the probability for a photon to produce a $\rho$ meson instead of converting into a $e^+e^-$ pair. When considering coherent production, $E_{\rho} \simeq E_\gamma$, and so we have 
\begin{equation}
    \frac{\dd N_\rho}{\dd E_\gamma}  = \frac{\dd N_\gamma}{\dd E_\gamma} \times \frac{\sigma_{\rm coh.}(E_\gamma)}{\sigma_{\rm pair}(E_\gamma)}\,.
\end{equation}
A similar formula holds for incoherent scattering, but the energies of the meson can be less than that of the incident photon, since the momentum transfer can be larger than in the coherent reaction. Following~\cite{Schuster:2021mlr}, we take $\sigma_{\rm coh.}(E_\gamma) = A f_{\rm coh.} \times \sigma_{\gamma p \rightarrow \rho p}(E_\gamma)$, where $\sigma_{\gamma p \rightarrow \rho p}$ is the photoproduction cross section on a free proton, and $A$ is the number of nucleons inside the nucleus. A similar formula defines the incoherent cross section, $\sigma_{\rm inc.}(E_\gamma) = A f_{\rm inc.} \times \sigma_{\gamma p \rightarrow \rho p}(E_\gamma)$. A detailed estimate including final-state interactions using a Glauber optical model for $A=27$ is given in~\cite{Schuster:2021mlr} from which we take $f_{\rm coh.}=0.57$ and $f_{\rm inc.}=0.35$ for $\rho$-meson production in all of our numerical estimates. 
We use a similar procedure as above to estimate $\phi$ meson photoproduction, with $f_{\rm coh.}=0.33$ and $f_{\rm inc.} = 0.65$ again following~\cite{Schuster:2021mlr}.  We use the smooth fit of $\rho$ and $\phi$ photoproduction from Fig.~1 of~\cite{Laget:2000gj} and demand $E_\gamma \geq 2~{\rm GeV}$.

The coherent reaction limits the momentum transfer to small values with a characteristic scale set by the inverse nuclear radius $1/R_{\rm Al} \simeq 55~{\rm MeV}$. This leads to small deflection angles that can be neglected when compared to the emission angle of $\chi$ or $\bar{\chi}$, which is set by the boost of the decaying meson, $\theta_{\chi} \sim m_\rho/E_\rho$. Incoherent scattering has larger momentum transfers ($|Q|\lesssim {\rm GeV}$); however, the mesons can still be treated as approximately forward-going (i.e., at small angles relative to the beam axis), since the decay angle dominates the final angle of the $\chi$ or $\bar{\chi}$ particle which later scatters in the detector.  We, therefore, treat all photoproduced $\rho$ and $\phi$ mesons as forward-going (i.e., we neglect their angle with respect to the beam axis when generating $\chi\bar{\chi}$ samples). Transverse momentum from the photoproduction of the parent meson is only important when the meson mass is close to $2m_\chi$, and we include this effect in our simulations. 

Given a sample of $\rho$ and $\phi$ mesons, we simulate their two-body decays into $\chi\bar{\chi}$ in the rest frame and boost the mCPs to the lab frame in order to get the total flux within the detector geometric acceptance (shown in \cref{fig:shower_production_number}). The contributions from meson decays peak near their kinematic thresholds after accounting for geometric acceptance effects, because the three-momentum of mCPs becomes small in the meson rest frame, making it easier to boost these mCPs into the forward region. In this near-threshold region, the transverse momentum distribution of the parent meson determines the height of the ``peaks'' that can be seen in \cref{fig:shower_production_number}. 

For $m_\chi \geq  m_\phi/2 \simeq 510~{\rm MeV}$, the decay of $\phi$ mesons is kinematically forbidden. The next relevant meson is the $J/\psi$. The larger meson mass, $m_{J/\psi} \approx 3.1~{\rm GeV}$, suppresses the coherent contribution to the cross section such that incoherent scattering (i.e., on individual nucleons) dominates the rate. Again following~\cite{Schuster:2021mlr} we simply take a naive model of incoherent scattering on $A$ independent nucleons at rest, since final-state interactions and other nuclear effects are expected to be small~\cite{Schuster:2021mlr}. The momentum transfers in the incoherent scattering are such that the energy of the $J/\psi$ differs substantially (i.e., by a few GeV) from the energy of the incident photon. 

To generate a sample of $J/\psi$ mesons we rely on recent data on $J/\psi$ production on a proton \cite{GlueX:2023pev} and demand $E_\gamma \geq 8.5~{\rm GeV}$. First, we draw photons from the weighted distribution
\begin{equation}
    \frac{\dd N_{\gamma \rightarrow J/\psi}}{\dd E_\gamma} = \frac{\dd N_\gamma}{\dd E_\gamma} \frac{A\times\sigma_{ J/\psi}(E_\gamma )}{\sigma_{\rm pair}(E_\gamma)}\,,
\end{equation}
where $\sigma_{ J/\psi}(E_\gamma)$ is the photoproduction cross section on a free proton \cite{GlueX:2023pev}. Given a fixed photon energy, we then sample $Q^2=|t|$ from the distribution $1/(1+Q^2/m_s^2)^4 \Theta(Q^2-Q^2_{\rm min}) \Theta(Q^2_{\rm max}-Q^2)$, where $m_s\approx 1.4~{\rm GeV}$ is fit from data \cite{GlueX:2023pev} (we take $m_s$ to be independent of energy for simplicity). The Heaviside functions enforce the minimal and maximal $Q^2$ given the kinematics. We find that the first generation of photons from $10^{22}~{\rm EOT}$ will produce roughly $6\times 10^{13}$ $J/\psi$ mesons in one radiation length. In the limit where $m_\chi \ll 1.5 ~{\rm GeV}$, this then leads to roughly $\varepsilon^2\times (5\times 10^{12})$ mCPs produced via $J/\psi \rightarrow \chi\bar{\chi}$.

The conversion from the $J/\psi$ Monte Carlo sample to a flux of $\chi$ and $\bar{\chi}$ passing through the detector is handled in the same manner as for $\rho$ and $\phi$ mesons. As shown in \cref{fig:shower_production_number}, the $J/\psi$ decay channel dominates the production rate when $m_\chi \gtrsim 500~{\rm MeV}$. This channel effectively determines the mass-reach of the electron beam dump setup we consider and there is a rapid loss in sensitivity for $m_\chi \gtrsim 1.55~{\rm GeV}$.

Our treatment of meson photoproduction is a conservative underestimate of the flux (in contrast to our treatment of the EM cascade production modes, which is realistic). We have only included photoproduction in the first radiation length, and we have neglected direct electroproduction (e.g., $e \ p \rightarrow e \ J/\psi \ p$). A more detailed simulation of photonuclear-induced meson decays at electron beam dump facilities will be pursued in future work, both for mCPs and, more generally, for dark sectors.

Finally, before moving onto mCP detection, let us comment on the propagation of mCPs through matter. Multiple Coulomb scattering of charged particles can lead to angular spreading which would influence the geometric acceptance plotted in \cref{fig:shower_production_number}. Between the aluminum target and the detector hall/well, there is 5.4~m of concrete and 14.2~m of dirt~\cite{Battaglieri:2020lds}. In \cref{app:MCS} we provide a short discussion of multiple Coulomb scattering, and conclude that it does  not significantly affect the angular distribution, and by proxy our sensitivity projections, for $\epsilon \lesssim 0.01$. We therefore do not include multiple Coulomb scattering in the intervening dirt and concrete between the target and the detector in any of the results that follow. 

\section{Detection with Low-Threshold Silicon Detectors}
\label{sec:detection}

Having discussed the production of mCP pairs in an electron beam-dump target, we now discuss how to detect them. We first discuss detecting mCPs using conventional (high-threshold) detectors, which have the advantage of being large, but the disadvantage of having higher thresholds, larger footprints, and being more expensive.  We then outline how low-threshold sensors can offer improved detection prospects for mCPs, focusing on silicon sensors.  We comment on backgrounds, such as beam-induced or cosmic-ray induced spallation neutrons, as well as cosmic-ray backgrounds more generally. 

\subsection{Conventional electron recoil detectors }\label{subsec:conventional}
\begin{figure}[t!]
    \centering
    \includegraphics[width=0.49\linewidth]{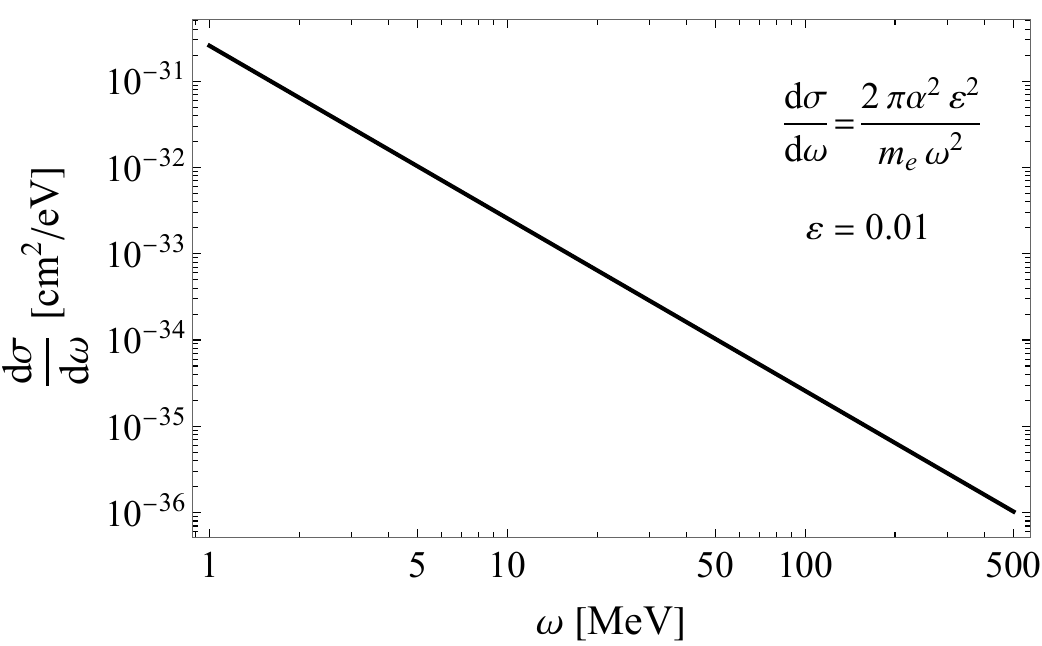}~
    \includegraphics[width=0.49\linewidth]{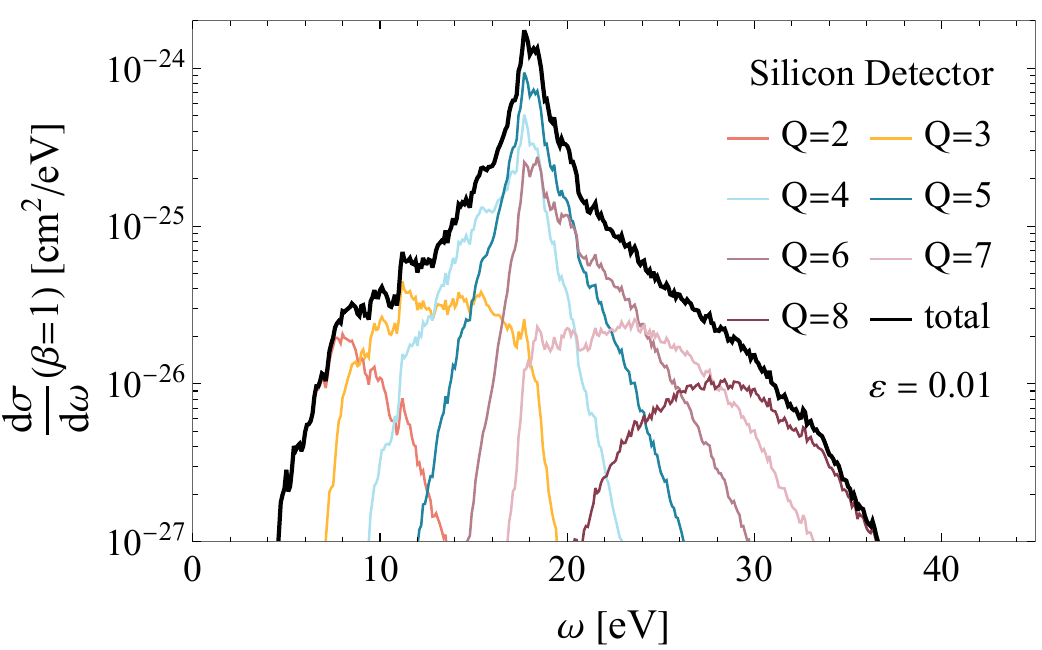}
    \caption{\textbf{Left:} Differential cross section per target electron for a relativistic mCP to scatter off a free electron ($\beta=1$). We set the millicharge to $\varepsilon=0.01$.  
    \textbf{Right:} Differential cross section per target electron for a relativistic mCP interacting with silicon ($\beta=1$). The black line shows the total cross section, while the colored lines show the cross section of producing an event with a specific charge $Q$, ranging from $Q=2e^-$ to $Q=8e^-$.
    }
    \label{fig:differential_cross_section}
\end{figure}

When electron recoil energies, $\omega$, are large compared to atomic scales ($\omega \gg  1~{\rm keV}$), we can approximate the electron as being free. The differential cross section for an mCP to scatter off a free electron is given by~\cite{Batell:2014mga,Harnik:2019zee}, 
\begin{equation}
     \left.\frac{\dd \sigma}{\dd \omega}\right|_{E_\chi\gg \omega,m_\chi,m_e}=\left.\pi\alpha^2\varepsilon^2\frac{2E_\chi^2m_e+\omega^2 m_e-\omega(m_\chi^2+m_e(2E_\chi+m_e))}{\omega^2(E_\chi^2-m_\chi^2)m_e^2}\right|_{E_\chi\gg \omega,m_\chi,m_e}\simeq\frac{2\pi\alpha^2\varepsilon^2}{\omega^2 m_e}\,.
\end{equation}
This differential cross section can be integrated to give the total cross section (denoted here as $\sigma_0$) for a recoil energy between $\omega^{\rm min}$ and $\omega^{\rm max}$~\cite{Harnik:2019zee}, 
\begin{equation}
    \!\!\!\sigma_0=\pi \alpha^2 \varepsilon^2 \frac {m_e(\omega^\mathrm{max}-\omega^\mathrm{min})(2 E_\chi^2 +\omega^\mathrm{max}\omega^\mathrm{min}) - \omega^\mathrm{max}\omega^\mathrm{min}\left(m_\chi^2 +m_e (2E_\chi+m_e)\right)\log\frac{\omega^\mathrm{max}}{\omega^\mathrm{min}}} {\omega^\mathrm{max}\omega^\mathrm{min} (E_\chi^2-m_\chi^2)m_e^2}\,.
\label{free-electron-recoil}
\end{equation}
Given the number of electrons in a detector, and the detection cross section, $\sigma_0$, one can compute the number of events per $10^{22}$~EOT. An important feature of \cref{free-electron-recoil} is that the cross section scales as $1/\omega^{\rm min}$, which leads to the well-known observation that low-threshold detectors have an increased sensitivity to mCPs. \cref{fig:differential_cross_section} (left) shows the differential scattering cross section versus the electron recoil energy $\omega$ assuming the free-electron approximation. In \cref{subsec:cross-section}, we will see that this is orders of magnitude below the peak of the scattering cross section (located near $\omega\sim 18$~eV) in a silicon detector (see \cref{fig:differential_cross_section}, right). 

The proposed BDX detector includes an electromagnetic calorimeter consisting of PbWO$_4$.  The planned location is 20~m downstream (and centered on the beamline) of the aluminum target with geometric dimensions given by $\sim 1~\text{m}\times1~\text{m}\times1~\text{m}$. For the BDX-mini experiment, which used PbWO$_4$ crystals totaling a volume of $\sim 4 \times 10^{-3}~{\rm m}^{-3}$, it was found that an energy threshold of $\sim 300~{\rm MeV}$ is required to achieve zero cosmic-ray backgrounds~\cite{BDX:2019afh,Battaglieri:2022dcy}.  Even if the 300~MeV energy threshold is sufficient to achieve zero backgrounds for the larger 1~m$^3$ detector, and despite its much larger geometric acceptance and detector mass, the mCP scattering cross section is very suppressed when compared to the peak cross section in a silicon detector, which lies below 50~eV.  We will illustrate the reach of such a setup (1~m$^3$ detector, 300~MeV threshold) in \cref{sec:results}, finding that it is much weaker than a small ($\mathcal{O}(10~{\rm cm}^2)$), gram-scale silicon detector with a few-eV threshold. 

Given the large increase in the scattering cross section for lower energy thresholds, it is interesting to ask at what threshold a hypothetical background-free 1~m$^3$ detector is comparable to a small, gram-scale low-threshold silicon detector.  We will see in \cref{sec:results}, that an energy threshold of about 10~MeV will allow for a comparable sensitivity provided that a zero-background search can be carried out with the same 10~MeV energy threshold.

\subsection{Silicon detectors with $\sim {\rm eV}$ thresholds}\label{subsec:cross-section}
Low-threshold sensors used for sub-GeV dark matter detection are also suitable for probing mCPs produced in colliding-beam, fixed-target, or beam-dump experiments. Although one might naively expect that the high energy of the mCP will also produce a high-energy signal in a detector, we saw in \cref{subsec:conventional} that the mCP scattering rate is highly suppressed at large recoil energies.  Instead, the signal peaks towards low energies. As we will review below, when a relativistic mCP scatters in a silicon detector target, it will typically produce events containing a few electron-hole pairs (which we will simply denote as ``$e^-$''); in fact, the majority of the signal events will contain a charge $Q$ of $4e^--6e^-$.  Low-threshold sensors, such as Skipper-CCDs~\cite{Tiffenberg:2017aac,SENSEI:2023gie} and other silicon or germanium detectors with TES or NTD readout~\cite{SuperCDMS:2018mne,SuperCDMS:2020ymb,SuperCDMS:2024yiv,EDELWEISS:2020fxc} have demonstrated single electron-hole pair resolution, and are capable of observing such signals.  

To calculate the cross section of relativistic mCPs in a sensor, we focus on silicon and use the formalism in~\cite{Essig:2024ebk}, which uses the dielectric function of a material. 
For a mCP particle with velocity $\beta$, the differential cross section (per electron) to deposit energy between $\omegaTrans$ and $\omegaTrans + \dd \omegaTrans$ in the target is given by
\begin{equation} \label{eqn:differential_cross_section}
    \frac{ \dd\sigma}{\dd \omegaTrans}(\omega,\varepsilon) = \frac{2 \alpha \varepsilon^2}{n_e \pi \beta^2} \int^{k_{\text{max}}}_{k_{\text{min}}} \\d k \left\{ \frac{1}{k} \text{Im}\qty( \frac{-1}{\epsilon(\omegaTrans, k)}) + k \qty ( \beta^2 - \frac{\omegaTrans^2}{k^2}) \text{Im}\qty(\frac{1}{-k^2 + \epsilon (\omegaTrans, k) \omegaTrans^2}) \right\}\,,
\end{equation}
where $\alpha$ is the fine structure constant, $n_e$ is the number density of free electrons in the target ($\sim$0.685~\AA$^{-3}$ for silicon), $k$ is the momentum transfer of the mCP to the target, and $\epsilon (\omegaTrans, k)$ is the dielectric function. The integration of the momentum transfer $k$ is kinematically constrained to be between $k_{\text{min}} = \omegaTrans/\beta$ and $k_{\text{max}} = 2|\mathbf{p}| - k_{\text{min}}$, where $\mathbf{p}$ is the momentum of the incoming mCP particle. The term $\text{Im}(-1/\epsilon(\omegaTrans, k))$
is the electron loss function (ELF), which describes the interaction of the mCP with the target through the exchange of a longitudinal-mode photon, which dominates in the context of non-relativistic scattering. In relativistic scattering, we also consider the exchange of a transverse-mode photon, which corresponds to the second term in the parentheses in \cref{eqn:differential_cross_section}; the term remains subdominant compared to the ELF in the range of $\omega$ that we are interested.
The mCPs arriving at the detector are all highly relativistic. For ease of computation, we simply evaluate the integral at $\beta=1$ in \cref{eqn:differential_cross_section}. 
For our results, we use an updated version of \texttt{QCDark}~\cite{Dreyer:2023ovn} to calculate the dielectric function~\cite{Dreyer:in-progress}. 

The cross section for producing $Q$ electrons is 
\begin{equation} \label{eqn:detection_cross_section}
\sigma_{Q}(\varepsilon) = 
\int^{\omegaTrans_{\text{max}}}_{\omegaTrans_{\text{min}}} \dd \omegaTrans\,P_Q(\omegaTrans)  \frac{\dd\sigma}{\dd \omegaTrans}(\omega,\varepsilon) \, ,
\end{equation}
where $\omegaTrans_{\text{min}}=1.1$ eV (the silicon band gap), $\omegaTrans_{\text{max}} = 50$ eV, and $P_Q(\omegaTrans)$ is the probability that a recoil energy $\omegaTrans$ produces an event with $Q$ electron-hole pairs.  For $P_Q(\omegaTrans)$, we use the results from~\cite{Ramanathan:2020fwm}.

We plot for a silicon sensor the differential cross section $\dd\sigma/\dd \omegaTrans$ from \cref{eqn:differential_cross_section} and also $\dd\sigma_Q/\dd \omegaTrans$ for various $Q$ in \cref{fig:differential_cross_section}, setting $\beta = 1$. 
We see a large peak around recoil energies of $\omega \sim 18$~eV, corresponding to $Q\sim 4e^- - 6e^-$.  The reason is that relativistic mCPs can excite collective modes (plasmons) in the silicon target, enhancing the rate significantly at the ``plasmon peak'', which is  at $\omegaTrans\sim18$~eV.  
As already discussed in~\cite{Essig:2024ebk}, the fact that most events contain multiple electrons is notable, since low-threshold sensors typically have larger backgrounds (e.g., from dark current) for $Q=1e^-$ and $2e^-$.  In particular, SENSEI has demonstrated background-free searches for $Q\sim 4e^- - 6e^-$~\cite{SENSEI:2023gie}.

\subsection{Background estimates \label{sec:backgrounds}}
As discussed in \cref{subsec:cross-section}, the mCP signal is distinct, peaking in the $Q=4e^--6e^-$ bins.  In general, backgrounds can do one of two things: they can produce events that mimic the signal, or they can require various analysis cuts that reduce the effective exposure. 
There are four types of backgrounds that could be a concern for the mCP search: beam-induced neutrons and muons, cosmic rays, radiogenic backgrounds, and pile up from single-electron events.  Whether these backgrounds are in fact a concern depends on the precise experimental setup and the specific sensors used in the readout. In particular, current Skipper-CCD sensors have very little timing information, so that it is impossible to perform an active veto of beam-induced or cosmic-ray induced backgrounds.  However, a TES- or NTD-based detector, for example, will have excellent timing resolution, so that an active veto is possible. On the other hand, the high spatial resolution of the Skipper-CCDs allow one to mask pixels with high charge density, focusing on those pixels that have no or very little charge.  Moreover, the proposed dual-sided CCD does have timing capabilities~\cite{Tiffenberg:2023vil}, and may be an excellent sensor for mCP searches and other searches at shallow underground sites.  We here simply motivate that a zero-background search seems plausible with low-threshold sensors after appropriate analysis cuts, but a careful study must of course be done by the proponents of a specific experiment. 

We find that beam-induced backgrounds are unlikely to be a concern, as long as the shielding after the dump is sufficiently extensive.  For example, the BDX collaboration has performed detailed analyses of the beam-related backgrounds in their scintillator detector. One possible concern is beam-induced neutrons, as they can be difficult to stop, but BDX finds that they are not a concern.  We revisit their arguments in \cref{app:backgrounds}, and also discuss why they are unlikely to be a concern also for the low-threshold sensors envisioned in our setup. 

Cosmic-ray and radiogenic backgrounds will typically deposit much larger amounts of charge than the mCPs.  For example, after masking high-energy events, the SENSEI search saw no events containing $Q=3e^--10e^-$ in 24~days of running at the $\sim$100~m underground MINOS cavern~\cite{SENSEI:2023gie}. However, the mCP search will have the sensor located at a much shallower underground site (for BDX, the detector would be located underground with an overburden of $\sim$10~meter-water-equivalent).  This site will have a much larger cosmic-ray muon backgrounds.  For low-threshold sensors with a good timing resolution, an active veto can help to remove neutrons induced from muons interacting near the detector.  For Skipper-CCDs, no active veto is possible.  However several experiments are already operating Skipper-CCDs near the surface or at very shallow underground sites (including very close to nuclear reactors) to search for coherent neutrino-nucleus scattering~\cite{CONNIE:2024pwt,CONNIE:2021ggh,CONNIE:2019swq,Depaoli:2024bgs}; moreover, recently the CONNIE and Atucha-II experiments performed a search for mCPs produced in reactor neutrinos~\cite{CONNIE:2024off}. 
While a background-free search seems plausible, the cosmic-ray and radiogenic backgrounds will at the very least reduce the effective exposure of the sensors to the mCP flux from the beam.  In our results below, we include a 20\% efficiency for this reduction in effective exposure. 

Finally, low-threshold sensors sometimes suffer from large single-electron-hole backgrounds, which can originate from detector-specific ``dark counts'' or from secondaries produced by cosmic-ray or radiogenic backgrounds~\cite{Du:2020ldo,Du:2023soy}. Pile-up from these single-electron events can mimic events containing multiple electron-hole pairs. Fortunately, the mCP signal peaks at larger charge bins, which are difficult to produce from the pile up of multiple single-electron-hole events.

\section{Results \label{sec:results}}
In the parameter region that we are interested in, the mCP cross section is small and we may treat the detector as a thin target. Our observable is the number of events in a fixed set of ``charge bins,'' which we take to be $Q\in \{3,4,5,6,7\}$. As mentioned above, due to cosmic-ray backgrounds, we assume a $20\%$ efficiency for a low-threshold sensor, $\epsilon_{\rm exp.}=0.2$, which accounts for the reduction in effective exposure. 
We believe this can plausibly lead to a background-free search, however a more detailed study of backgrounds will be necessary. We note that the projected sensitivity to the millicharge $\varepsilon$ scales as $\epsilon_{\rm exp.}^{1/4}$.

\begin{figure}[t!]
    \centering
    \centering
    \includegraphics[width=0.45\linewidth]{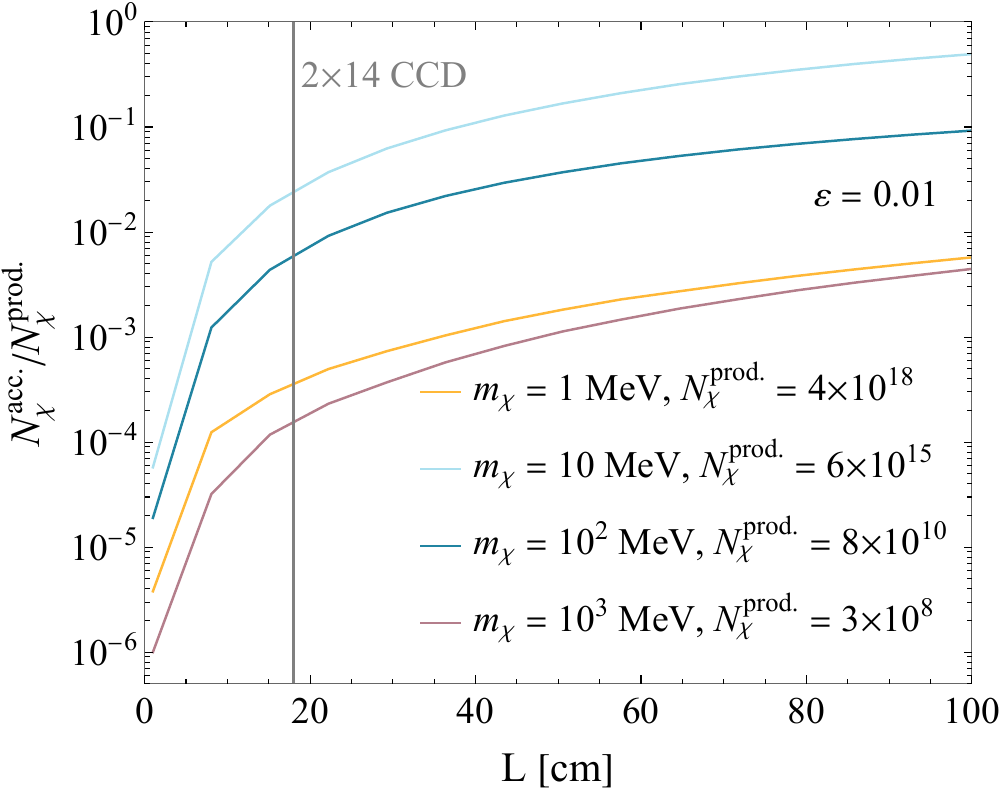}~~ \hfill
    \includegraphics[width=0.495\linewidth]{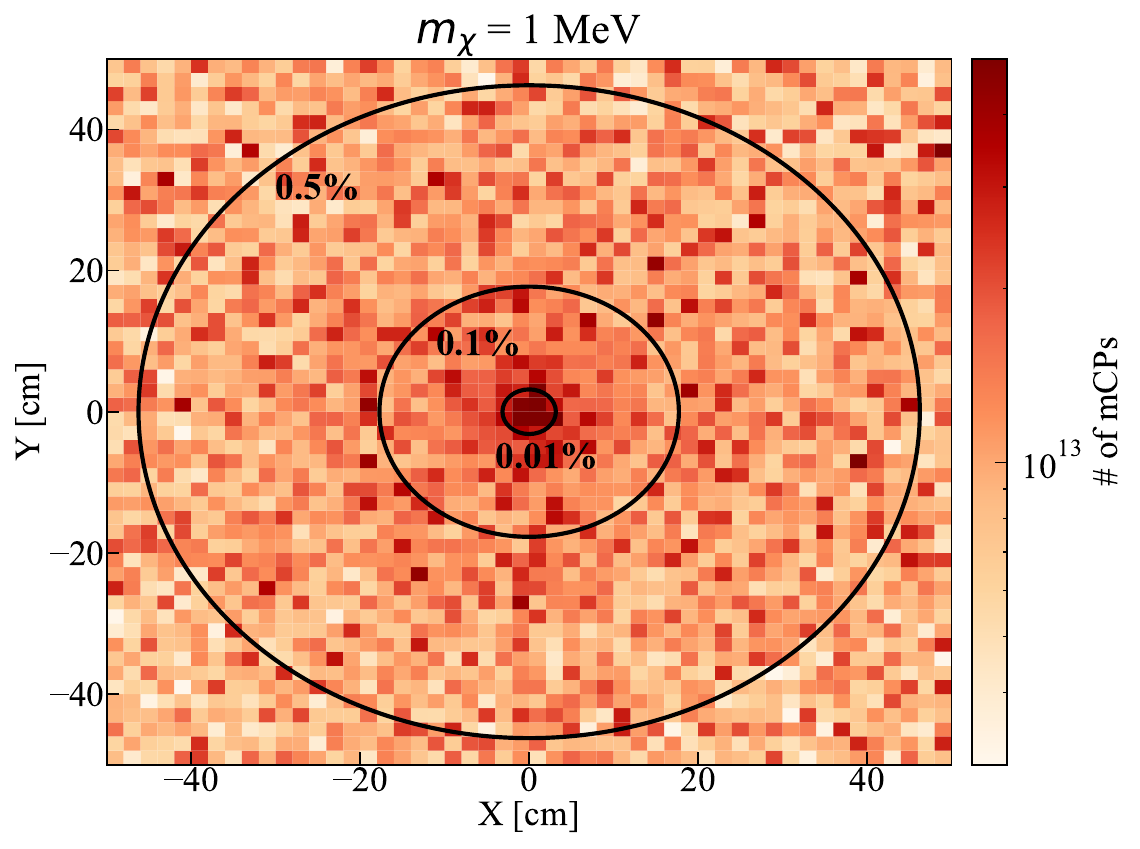}
    \\
    \includegraphics[width=0.495\linewidth]{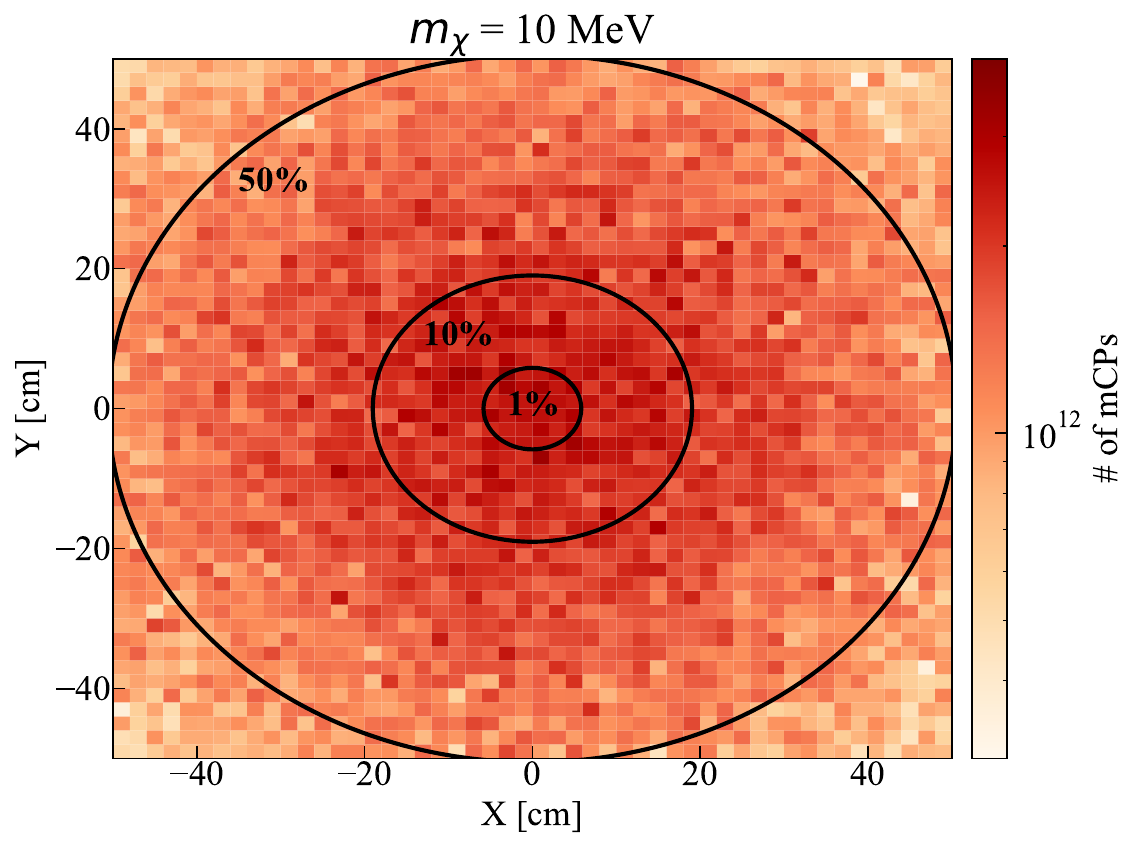}
    \includegraphics[width=0.495\linewidth]{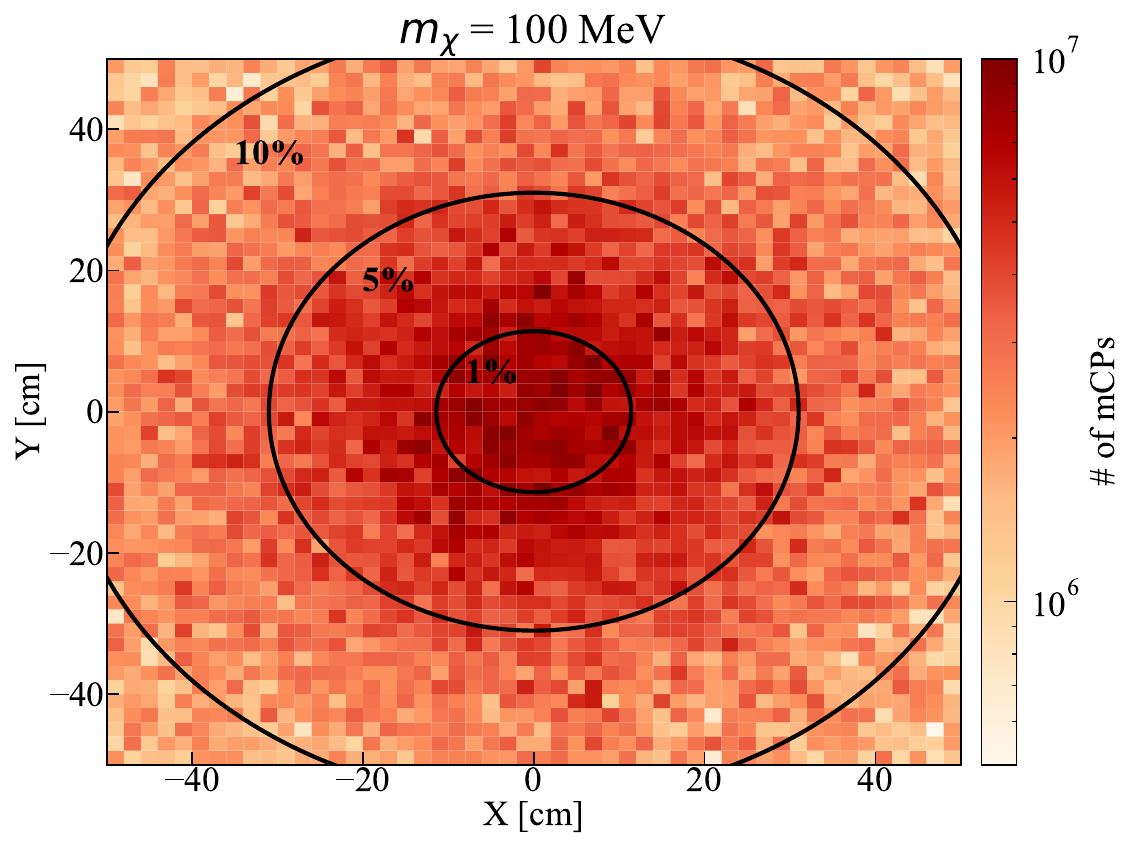}
    \caption{
    \textbf{Top left:} The geometric acceptance for a square detector of side length $L$ [cm] for different mCP masses (in MeV). The curves and contours are normalized by $N^\text{prod.}_\chi$ as given in the legend. The scaling of the acceptance with particle mass can be understood by considering the different production mechanisms. The geometric acceptance from mCP annihilation increases with $m_\chi$ because of a decrease in transverse momentum. At higher masses, acceptance is determined by the kinematics of the parent meson, which are only moderately boosted leading to the much lower acceptance for $m_\chi=1~{\rm GeV}$. 
    \textbf{Top right:} A color map that shows the number of signal events passing the X-Y plane at which the detector is located, for $\varepsilon=0.01$ and $m_\chi = 1$~MeV. The contours show the accepted events as a fraction of the total produced.  
    \textbf{Bottom left} and \textbf{bottom right} are as the top right figure, but for mCP masses $m_\chi = 10$~MeV and $m_\chi = 100$~MeV, respectively. 
    } 
    \label{fig:geometric_acceptance}
\end{figure}

The expected number of signal events can therefore be written as 
\begin{equation}\label{eq:signal-number}
    N_\text{detected}(\varepsilon, m_{\chi}) = \epsilon_{\text{exp.}} \times N^\text{acc.}_\chi (\varepsilon, m_\chi) \times (n_e  L) \times  \sum^7_{Q=3} \sigma_{\text{Q}} (\varepsilon) \: ,
\end{equation}
where $\epsilon_{\rm exp.}=0.2$ is the detector efficiency mentioned above, $L=665~\mu{\rm m}$ is the thickness of the detector, and $n_e=6.84\times10^{23}~{\rm cm}^{-3}$ is the number density of electrons in silicon. The accepted number of mCPs, $N^\text{acc.}_\chi$, includes the geometric acceptance of the detector located 20~m downstream of the target on the beam line. In \cref{fig:geometric_acceptance}, we show the geometric efficiency for various detector areas and mCP masses. 
The cross section, $\sigma_Q$, per bin in $Q$ is given in \cref{eqn:detection_cross_section} (see also \cref{fig:differential_cross_section}).

In the remainder of this section, we consider two mCP scenarios (a) pure millicharged particle and (b) millicharged DM that interacts with a hidden $U(1)$ gauge boson, which is kinetically mixed with ordinary photon, and discuss relevant constraints for each model. We provide the details for each model separately, including a discussion of existing constraints from the literature. 

\vfill 
\pagebreak

\subsection{Pure millicharged particles}

\begin{figure}[!t]
    \centering
    \includegraphics[width=0.495\linewidth]{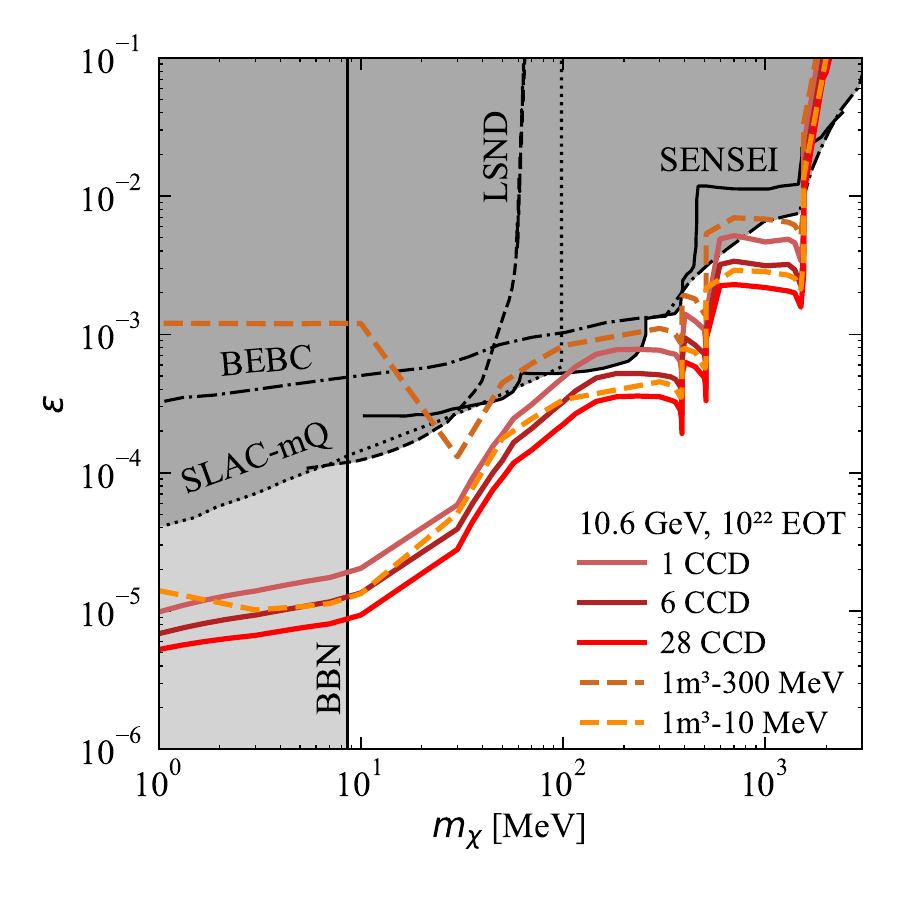}
    \includegraphics[width=0.495\linewidth]{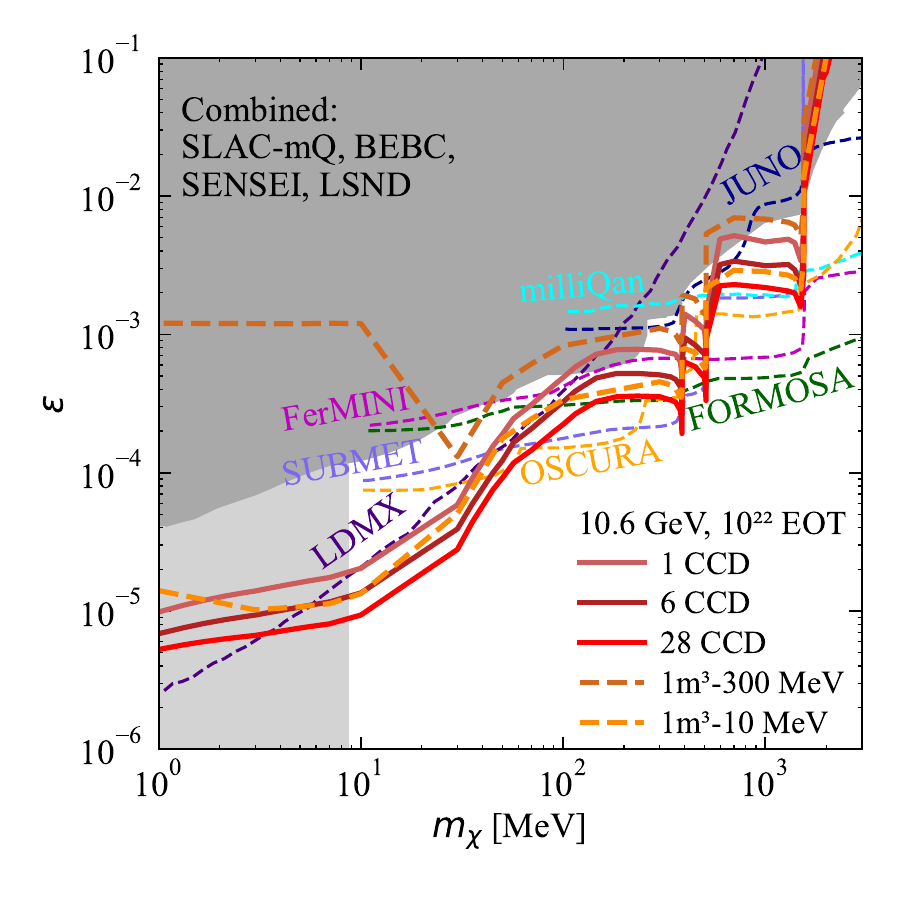}
    \caption{
    \textbf{Pure mCP model.} \textbf{Left:} Projected sensitivity to mCPs in the millicharge $\varepsilon$ versus mCP mass $m_\chi$ plane of a Skipper-CCD detector array placed near the proposed BDX detector at Jefferson Laboratory.  We assume $10^{22}$ electrons on target and show the sensitivity for three different low-threshold silicon detector choices (lines with various shades of red): a single $9.2 \times 1.3$~cm$^2$ Skipper-CCD, 6 Skipper-CCDs ($9.2~\times 7.8$~cm$^2$), and a larger square detector array ($\sim 18\times 18~{\rm cm}^2$, corresponding to 28-Skipper-CCDs). We also show two sensitivities of a hypothetical lead tungstate detector (electron number density of $n_e = 2.06 \times 10^{24}$~cm$^3$) with volume 1~m$^3$ and an electron-recoil energy threshold of 10~MeV and 300~MeV (two different dashed orange lines), both assumed to be background-free. The most competitive collider constraints are shown in dark gray, and the $N_{\text{eff}}$ bound from BBN is shown in light gray. \textbf{Right:} Same as left plot, but here we show other proposed experiments in dashed curves. We call particular attention to milliQan~\cite{milliQan:2021lne} in cyan,  and SUBMET~\cite{Kim:2021eix} in lavender (which is currently taking data).  Other proposed accelerator-based experiments include LDMX~\cite{Berlin:2018bsc} in purple, Fermini in magenta~\cite{Kelly:2018brz}, FORMOSA in green~\cite{Foroughi-Abari:2020qar}. We also show the projected sensitivity of the Oscura detector~\cite{Oscura:2023qch}. The estimated sensitivity for a proposed search for cosmic ray produced mCPs in the JUNO detector~\cite{ArguellesDelgado:2021lek,Wu:2024iqm} is shown in blue.
    \label{fig:pure_mCP_plot}}
\end{figure}

We first consider a ``pure'' mCP i.e., without any additional gauge force carrier. The simplest extension of the Standard Model of this form include a new fermion with a small hypercharge $Y=\varepsilon e$, which transforms as a singlet under $SU(2)$. This leads to a small coupling to the $Z$ boson as well as the photon; however, constraints on the $Z$-coupling are always much weaker than the coupling  to photons.  For simplicity in what follows, we always consider a Dirac fermion, however our results are applicable up to $O(1)$ factors to any particle with a small electric charge. 

In \cref{fig:pure_mCP_plot}, we show the projected sensitivity of a Skipper-CCD detector (array) placed near the proposed BDX experiment at JLab. We include the sensitivity projections for using 1, 6, and 28 Skipper-CCD(s). These choices correspond to the minimal setup (a single $9.2 \times 1.3$~cm$^2$ Skipper-CCD), a detector array that could be readily deployed using existing detectors (6 Skipper-CCDs), and a larger square detector array ($\sim 18\times 18~{\rm cm}^2$, corresponding to 28-Skipper-CCDs), respectively. We include constraints from the electron beam dump experiment at SLAC (SLAC-mQ)~\cite{Prinz:1998ua}, the LSND experiment \cite{LSND:2001akn,Magill:2018tbb}, and proton beam dump experiments at CERN (BEBC) \cite{Marocco:2020dqu} and Fermilab (SENSEI) \cite{SENSEI:2023gie}. We also show constraints from  $N_{\text{eff}}$ that are obtained from an analysis of mCPs 
 during Big Bang nucleosynthesis (BBN) \cite{Boehm:2013jpa, Creque-Sarbinowski:2019mcm}. The curve for the 1~${\rm m}^3$ detector (with a 300~${\rm MeV}$ threshold) shown in \cref{fig:pure_mCP_plot} has a sharp ``upturn'' around $m_\chi=20$~MeV. This occurs because, as $m_\chi$ decreases, a larger fraction of mCPs have either low $E_\chi$ or high transverse momentum. These two variables are anti-correlated in such a way that most of the flux that hits the $1~{\rm m}^3$ detector falls below the $300~{\rm MeV}$ threshold. A similar, albeit less stark, effect can be seen for the $10~{\rm MeV}$ threshold curve, while the effect is non-existent for the Skipper-CCD due to its incredibly small energy threshold.

In the right-panel of \cref{fig:pure_mCP_plot}, we include projections from proposed experiments in the literature. Of particular note is the proton beam experiment at J-PARC (SUBMET)~\cite{Kim:2021eix}, which is currently taking data, and the milliQan experiment at CERN, whose demonstrator has already produced results~\cite{milliQan:2021lne}. We also show projected sensitivity from a missing momentum search using LDMX at SLAC~\cite{Berlin:2018bsc}, a proposed setup using a milliQan-like detector in the Fermilab NuMI beam (FerMINI)~\cite{Kelly:2018brz}, a proposed setup using a similar detector in the a proposed forward physics facility at the LHC (FORMOSA)~\cite{Foroughi-Abari:2020qar}, and a proposed setup using a Skipper-CCD detector (Oscura) in the NuMI beam line~\cite{Oscura:2023qch}.  Finally, beyond accelerator facilities, we also include projections from a mCP search at the Jiangmen Underground Neutrino Observatory (JUNO)~\cite{Wu:2024iqm} (see also Ref.~\cite{ArguellesDelgado:2021lek}) using the mCP flux that can be produced by cosmic rays showers in the upper atmosphere \cite{Plestid:2020kdm,ArguellesDelgado:2021lek,Du:2022hms,Wu:2024iqm}.

The sensitivity with a Skipper-CCD detector array in a BDX-like beam dump is competitive with essentially all proposed experiments in the $\sim 10~{\rm MeV}-1~{\rm GeV}$ mass range. Our results suggest that world-leading sensitivity can be obtain for $m_\chi \lesssim 70~{\rm MeV}$, and that the electron beam setup will be competitive with the SUBMET search over their full mass-reach window. The sensitivity at low masses is dominated by the mCP flux from annihilation and trident production. Above $\sim 130~{\rm MeV}$ the flux from vector meson decays begins to dominate (the same features can be seen in the predicted SUBMET sensitivity curves).

\subsection{Dark matter interacting with an ultralight dark photon}

In the context of mCPs, it is interesting to explore the complementarity of accelerator-based probes of dark sectors and the direct detection of dark matter. 
We consider dark matter that is coupled to an ultralight dark photon that is kinetically mixed~\cite{Holdom:1985ag,Galison:1983pa} with the ordinary photon.  Specifically, we will focus on the following low-energy effective Lagrangian 
\begin{equation}
    \mathcal{L} \supset -\frac14\mathcal{F}_{\mu\nu}\mathcal{F}^{\mu\nu} + \frac12 m_{\mathcal{A}}^2 \mathcal{A}_\mu \mathcal{A}^\mu -\frac{\kappa}{2}F_{\mu\nu}\mathcal{F}^{\mu\nu} + A_\mu J^\mu -g_D \mathcal{A}_\mu\bar{\chi}\gamma^\mu \chi  ~, 
\end{equation}
where $J_\mu$ is the Standard Model electromagnetic current, $F_{\mu\nu}$ ($\mathcal{F}_{\mu\nu}$) is the field strength of the ordinary (dark) photon $A_\mu$ ($\mathcal{A}_\mu$), and $\kappa$ the kinetic mixing parameter. The mass $m_{\mathcal{A}}$ can either arise from the Stuckelberg or Higgs mechanisms. For $m_{\mathcal{A}} \ll 1~{\rm eV}$, $\chi$ is very similar to an mCP.  After rotating to the mass basis, the dark photon couples to $J^\mu$ (and therefore the electron bilinear $\bar{\psi}_e\gamma^\mu \psi_e$) with a coupling strength of $\varepsilon e$, with    
\begin{equation}
    \varepsilon = \kappa g_D ~,
\end{equation} 
acting as an ``effective'' millicharge (even though $\chi$ does not actually have an electric charge for non-zero $m_{\mathcal{A}}$).  In particular, all bounds on mCPs also apply to the $\chi$ particles. 

Direct-detection searches for $\chi$ with low-threshold detectors have a distinct advantage over accelerator probes, since the scattering cross section scales as $1/q^4$, where $q$ is the momentum transfer.  Very small couplings can be probed in the future~\cite{Essig:2011nj,Essig:2015cda}, even those required to produce the correct relic abundance from freeze-in~\cite{Essig:2011nj,Hall:2009bx,Chu:2011be,Essig:2015cda,Dvorkin:2019zdi}. Current direct-detection searches already rule out a large range of couplings above the values required for freeze-in.  However, for sufficiently large values, the dark-matter particles $\chi$ interact so strongly with ordinary matter that they would get stopped in the Earth's atmosphere and crust, so that there are no direct-detection constraints~\cite{Emken:2019tni}.  For these large values, however, strong bounds from the Cosmic Microwave Background (CMB) limit the amount of dark matter to $\lesssim$ 0.4\%~\cite{Boddy:2018wzy}, thus restricting the $\chi$ to be a subdominant component of dark matter.  It is interesting that the allowed values of $\varepsilon$ are sufficiently large that they can be (partially) probed by accelerator-based experiments. 

We note that pure mCPs may constitute the dark matter, but in the parameter region of interest for accelerator-based probes, if the dark matter were pure mCP, then (given its non-relativistic\footnote{We note that a pure mCP dark matter subcomponent will generate a flux with relativistic velocities from Rutherford scattering with cosmic rays, and that this flux can be enhanced by magnetic retention. This can lead to interesting and complimentary direct detection prospects \cite{Harnik:2020ugb}.} virial velocity) it would  not be able to penetrate the solar wind to intersect a terrestrial detector, and there are thus no direct-detection constraints~\cite{Dunsky:2018mqs,Emken:2019tni}.  Instead, the presence of a massive dark photon screens the magnetic fields on length scales $\gg 1/m_\mathcal{A}$, allowing the $\chi$ to penetrate the solar wind for appropriate choices of $m_\mathcal{A}$. We thus focus on dark matter particles that interact with a massive, albeit ultralight, dark photon, for which there is an exciting complementarity between direct-detection and accelerator-based probes.

\begin{figure}[!t]
    \centering
    \includegraphics[width=0.495\linewidth]{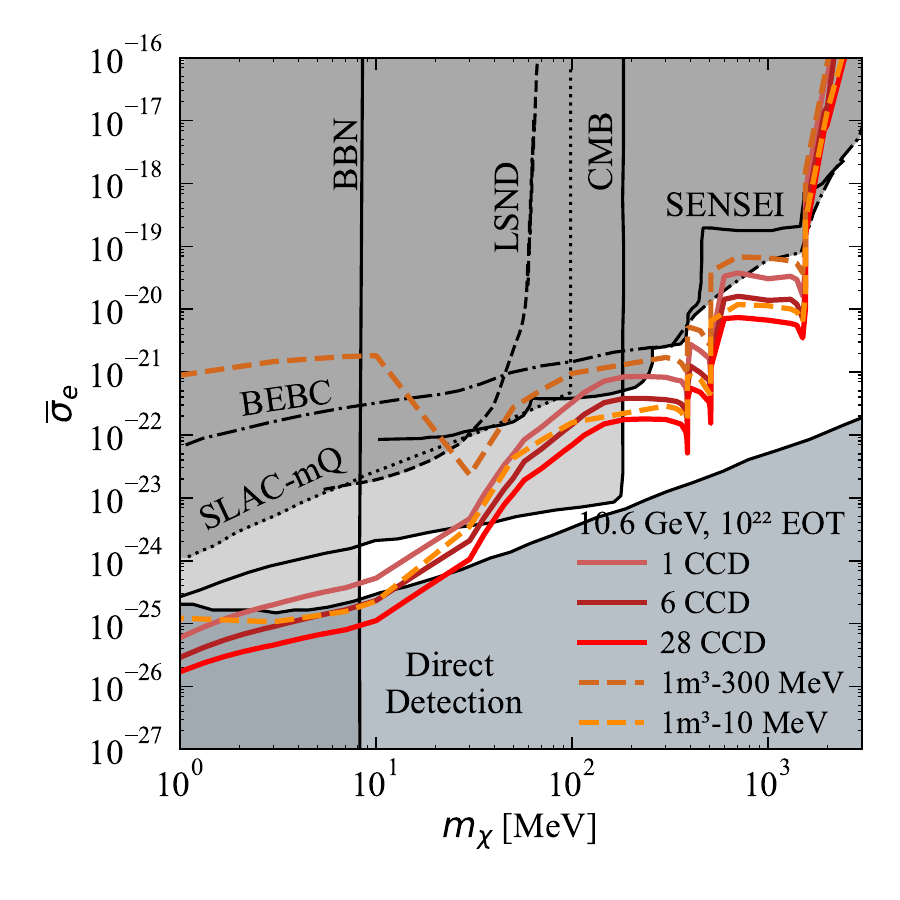}
    \includegraphics[width=0.495\linewidth]{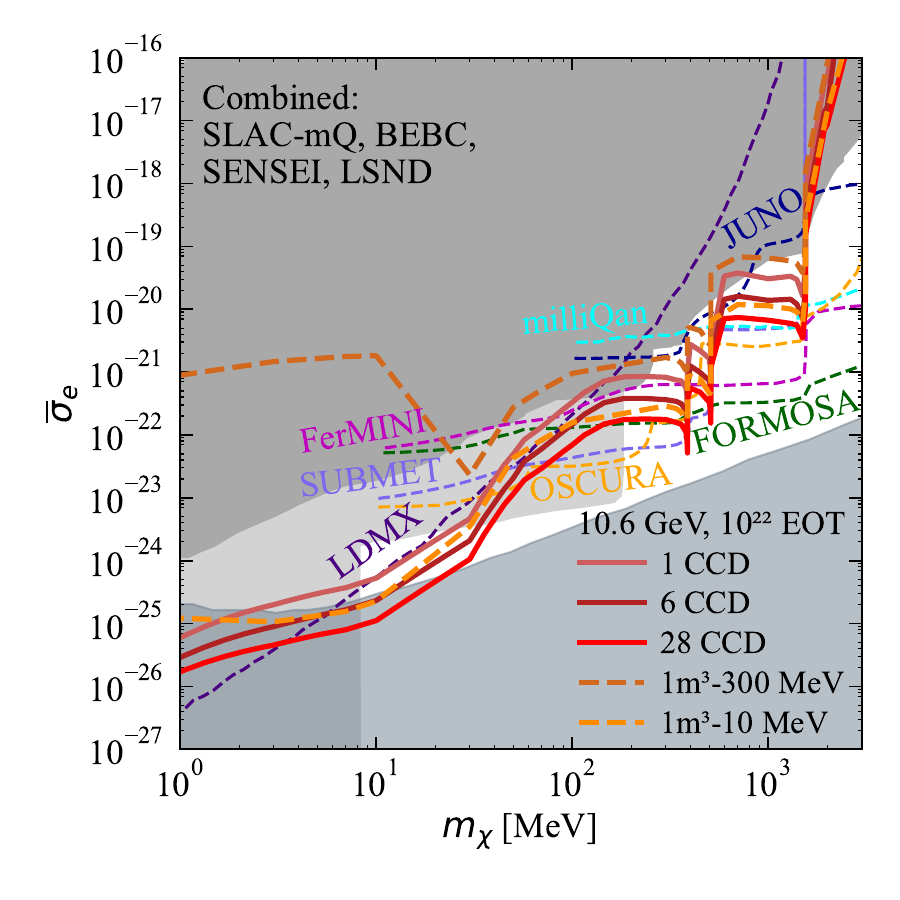}
    \caption{
    \textbf{Dark matter interacting with an ultralight dark photon.} \textbf{Left:} Projected sensitivity to dark matter interacting with an ultralight dark photon in the $\overline{\sigma}_e$ versus dark matter mass ($m_\chi$) plane of a Skipper-CCD detector array placed near the proposed BDX detector at Jefferson Laboratory.  We assume $10^{22}$ electrons on target and show the sensitivity for three different detector array choices (lines with various shades of red): a single $9.2 \times 1.3$~cm$^2$ Skipper-CCD, 6 Skipper-CCDs ($9.2~\times 7.8$~cm$^2$), and a larger square detector array ($\sim 18\times 18~{\rm cm}^2$, corresponding to 28-Skipper-CCDs). We also show two sensitivities of a hypothetical lead tungstate detector (electron number density of $n_e = 2.06 \times 10^{24}$~cm$^3$) with volume 1~m$^3$ and an electron-recoil energy threshold of 10~MeV and 300~MeV (two different dashed orange lines), both assumed to be background-free. The collider constraints in dark gray and the BBN constraints in light gray are the same as in \cref{fig:pure_mCP_plot}. The direct-detection constraints (combining all limits) are included in gray-blue~\cite{Emken:2019tni}. We assume a dark gauge coupling of $g_D=0.1$ and a dark-matter fractional abundance of less than 0.4\%. 
    \textbf{Right:} Same as left plot, but here we show other proposed experiments in dashed curves, as in \cref{fig:pure_mCP_plot}. 
    \label{fig:DM_plot}}
\end{figure}

Accelerator probes are best understood in the plane of $\varepsilon$ and $m_\chi$ shown in \cref{fig:pure_mCP_plot}. These constraints can be mapped onto the direct-detection cross-section versus dark-matter mass plane,  by making use of a (momentum-independent) ``reference cross section'', $\overline{\sigma}_e$, defined as~\cite{Essig:2011nj,Essig:2015cda}
\begin{equation}
    \overline{\sigma}_e = \frac{16 \pi \alpha^2\varepsilon^2 \mu_{\chi e}^2}{(\alpha m_e)^4}\,,
\end{equation}
where we have assumed $m_\mathcal{A} \ll \alpha m_e$, and where $\mu_{\chi e}$ is the reduced mass of DM and electron. The actual scattering rate is proportional to $\overline{\sigma}_e \times F^2_{\rm DM}(q)$, where the dark-matter form factor $(F_{\rm DM}=(\alpha m_e/q)^2$ captures the momentum-dependence of the scattering. 

In  \cref{fig:DM_plot}, we include all of the previous pure mCP constraints and projections, now in terms of the reference cross section $\overline{\sigma}_e$.  We assume the $\chi$'s constitute much less than 0.4\% of the dark-matter abundance, which avoids the very strong CMB constraints that arise from the $\chi$ scattering off ordinary matter in the early Universe~\cite{Boddy:2018wzy}.   We also include combined direct detection constraints from SENSEI, CDMS-HVeV, XENON10, XENON100, DarkSide-50 experiments~\cite{XENON:2016jmt,Essig:2017kqs,DarkSide:2018ppu,Crisler:2018gci,SuperCDMS:2018mne,Laletin:2019qca} with the compiled limits take from~\cite{Emken:2019tni}, whose upper boundary is insensitive to the precise value of the dark matter abundance.  We do not show bounds that are sensitive to the precise abundance, e.g.~\cite{Rich:1987st,Mahdawi:2018euy,Prabhu:2022dtm}. These constraints are superimposed on the accelerator and BBN constraints discussed in \cref{fig:pure_mCP_plot}, which are applicable independent of the model assumptions related to, e.g., the fraction of dark matter or the mass of the light-mediator.  In addition, the presence of a low-mass dark photon in the model is constrained by the CMB bounds on the effective number of relativistic degrees of freedom, $N_{\text{eff}}$; we choose the most conservative bound, which arises for relatively large values of the dark gauge coupling (e.g., $g_D=0.1$)~\cite{Creque-Sarbinowski:2019mcm, Emken:2019tni, Munoz:2018pzp, Vogel:2013raa}. We see that the BDX-like setup we study in this paper is capable of probing a large fraction of the open window.


%
\section{Discussion and Conclusions \label{sec:conclusions}}%

In this work, we have demonstrated that a high-intensity, medium-energy electron beam dump together with an ultralow-threshold sensor offers world-leading sensitivity to mCPs. Searching for low-mass dark sectors at accelerator facilities is now widely appreciated as a complimentary method by which to probe a variety of dark-sector models that have interesting and non-trivial cosmological histories. Models with a mCP in their spectrum represent a minimal example of such a sceanrio and are unique in that their detection signature benefits from ultralow thresholds. Since these ultralow threshold detectors are small (e.g., gram-scale) they can be easily deployed.  

In our paper, for concreteness, we have performed a detailed study of mCP production in the aluminum target planned for the Beam Dump eXperiment (BDX) at J-Lab. We find that a number of previously overlooked production modes (namely $e^+e^-$ annihilation and vector-meson photoproduction) substantially increase the sensitivity of electron beam dumps to mCPs and likely to dark sectors more generally. Our findings suggest that the SLAC-mQ exclusions may be underestimated, and suggest that a reanalysis of their model for mCP production in the target is worth revisiting.

Looking forward, our study further motivates the inclusion of small ultralow-threshold sensors in future beam-dump and collider facilities. Although we have focussed on electron beam dumps a small Skipper-CCD array like the ones discussed above could be easily included in facilities such as SHiP, the LHC's forward physics facility, or in the DUNE near-detector complex. We encourage further exploration of ultralow-threshold detectors in beam-dump experiments~\cite{LDMX:2018cma,Balasubramanian:2023pap,NA64:2023wbi,Jena:2024ycp,SHiP:2021nfo} and in the forward region of colliders such as the LHC~\cite{FASER:2019aik,Graverini:2024ynx,milliQan:2021lne,Batell:2021blf,Foroughi-Abari:2020qar,Feng:2022inv}.  

\acknowledgments

We thank Duncan Adams for helpful conversations related to neutron backgrounds, Marco Battaglieri for helpful correspondence about the BDX detector, Christopher Hill for correspondences on the SUBMET experiment, Javier Tiffenberg for useful comments on CCD readout and cosmic-ray induced backgrounds, and Kevin Zhou for helpful discussions about vector meson photoproduction. 
RE acknowledges support from DOE Grant DE-SC0025309, Simons Investigator in Physics Award~623940, Heising-Simons Foundation Grant No.~79921, Binational Science Foundation Grant No.~2020220.  
MD and HX are supported in part by DOE Grant DE-SC0009854 and Simons Investigator in Physics Award~623940.  In addition, HX is supported in part by the Binational Science Foundation Grant No.~2020220.
PL and ZL are supported in part by the DOE Grant No.~DE-SC0011842 and a Sloan Research Fellowship from the Alfred P.~Sloan Foundation at the University of Minnesota. RP is supported by the Neutrino Theory Network under Award Number DEAC02-07CH11359, the U.S.~Department of Energy, Office of Science, Office of High Energy Physics under Award Number DE-SC0011632, and by the Walter Burke Institute for Theoretical Physics.

\appendix

\section{Dependence of electromagnetic shower simulation on energy threshold \label{app:energy-thesholds}}
When simulating particle production in an electromagnetic cascade one must supply a prescription for when to halt the numerical simulation of the cascade. In \texttt{PETITE} this is done using a minimum energy threshold. In this appendix, we show how the predicted fluxes arriving at the Skipper-CCD converge as a function of the minimum energy threshold. We will explicitly see that our results are not sensitive to variations in the minimum energy threshold. 

\subsection{Trident production}
\begin{figure}[!t]
    \centering
    \includegraphics[width=0.49\textwidth]{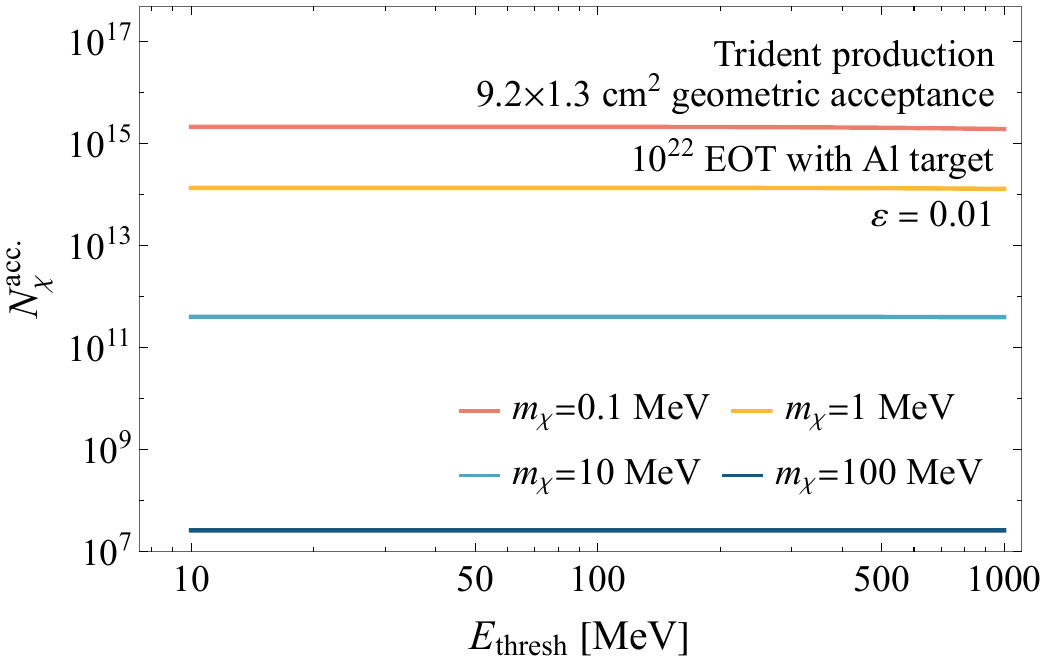}~
    \includegraphics[width=0.49\textwidth]{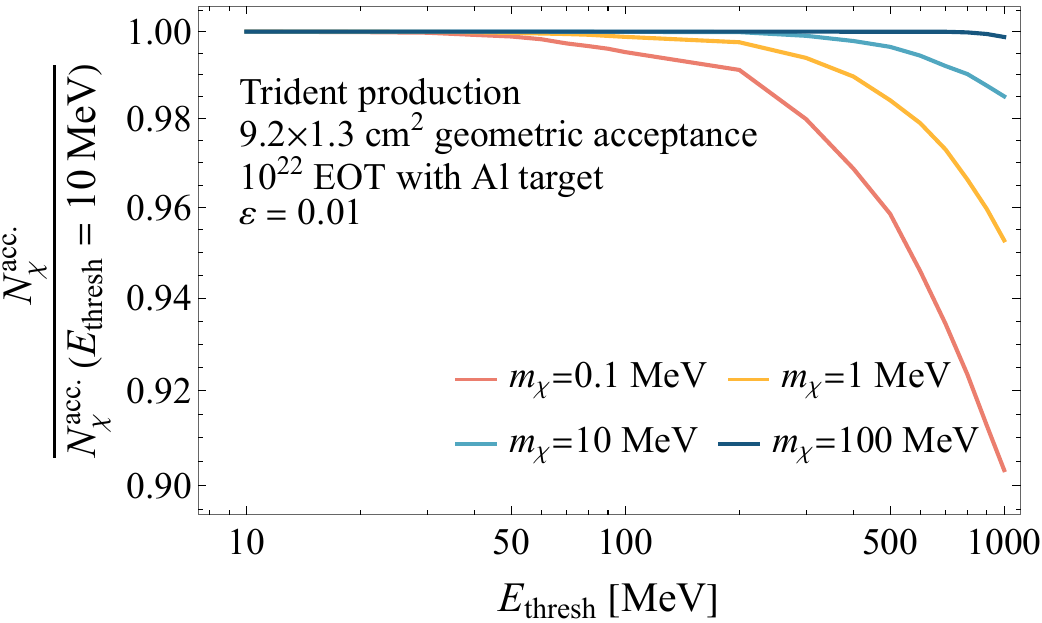}
    \caption{
    \textbf{Left:} The number of mCPs produced in trident processes in the EM shower that pass through a detector of area $9.2 \times 1.3~{\rm cm}^2$ located 20~m downstream of the target ($N_\chi^\text{acc.}$) as a function of the shower energy threshold, for a millicharge of $\varepsilon=0.01$.  
    \textbf{Right:}
    The ratio of $N_\chi^\text{acc.}$ to    $N_\chi^\text{acc.}(E_{\rm thresh} = 10\text{MeV})$, the accepted number of mCPs evaluated at a shower energy threshold of 10~MeV. We see that for $E_{\rm thresh}\lesssim 50~{\rm MeV}$ the flux hitting the detector is insensitive to the shower threshold.
    \label{fig:trident_shower_number}}
\end{figure}

In \cref{fig:trident_shower_number} (left), we show the number of mCPs produced in trident processes in the EM shower that pass through a detector of area $9.2~{\rm cm}\times 1.3~{\rm cm}$ located 20~m downstream of the target ($N_\chi^\text{acc.}$) versus the energy threshold.  In the right plot, we show the ratio of $N_\chi^\text{acc.}$ to the accepted number of mCP evaluated at a shower energy threshold of 10~MeV, denoted $N_\chi^\text{acc.}(E_{\rm thresh} = 10~\text{MeV})$. For low-mass mCPs (e.g., $m_\chi=0.1$~MeV), the accepted number of mCPs from trident production increases by about 10\% from lowering the energy threshold from 1~GeV to 50~MeV, and remains flat for even lower energy threshold values. Although the number of low-mass mCPs produced increases for lower energy thresholds, most of them will not go through the small downstream detector area.  
For heavier mCPs, such as for $m_\chi = 100$~MeV, an energy threshold of 1~GeV is already sufficiently low to properly estimate the number of mCPs that pass through the detector. The reason is that the majority of secondary particles do not have enough energy to produce heavy mCPs, and those that do, will typically produce them at large angles.  
In any case, we see here that an energy threshold of $E_\text{thresh}= 10~{\rm MeV}$ is more than sufficient for robustly estimating the number of mCPs produced through trident processes in the EM shower. 

\subsection{Annihilation production}
\begin{figure}[!t]
    \centering
    \includegraphics[width=0.49\textwidth]{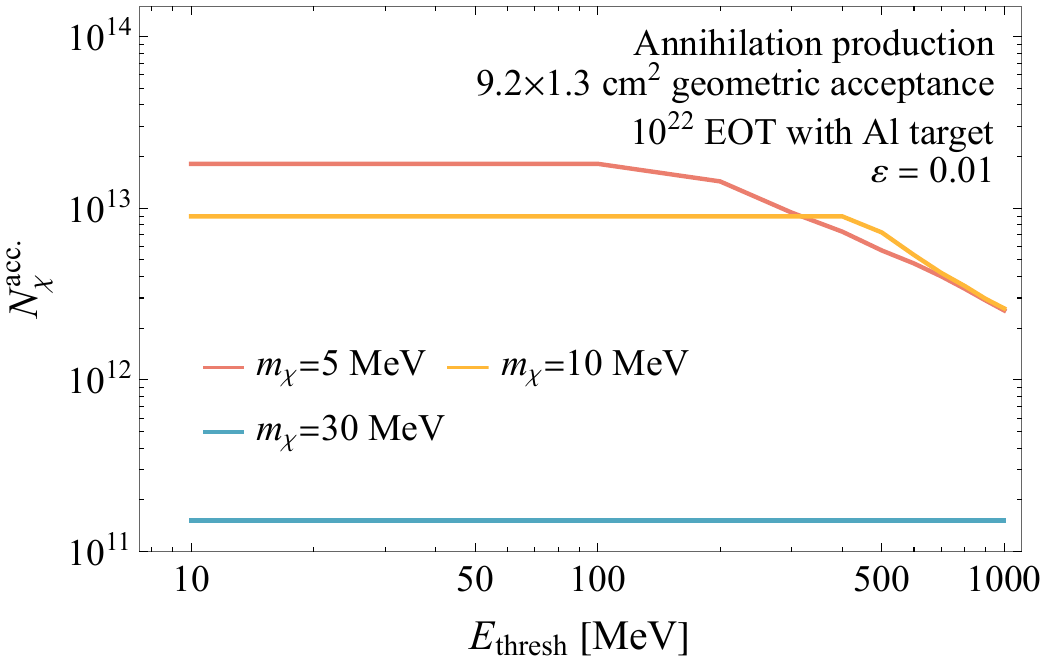}~
    \includegraphics[width=0.49\textwidth]{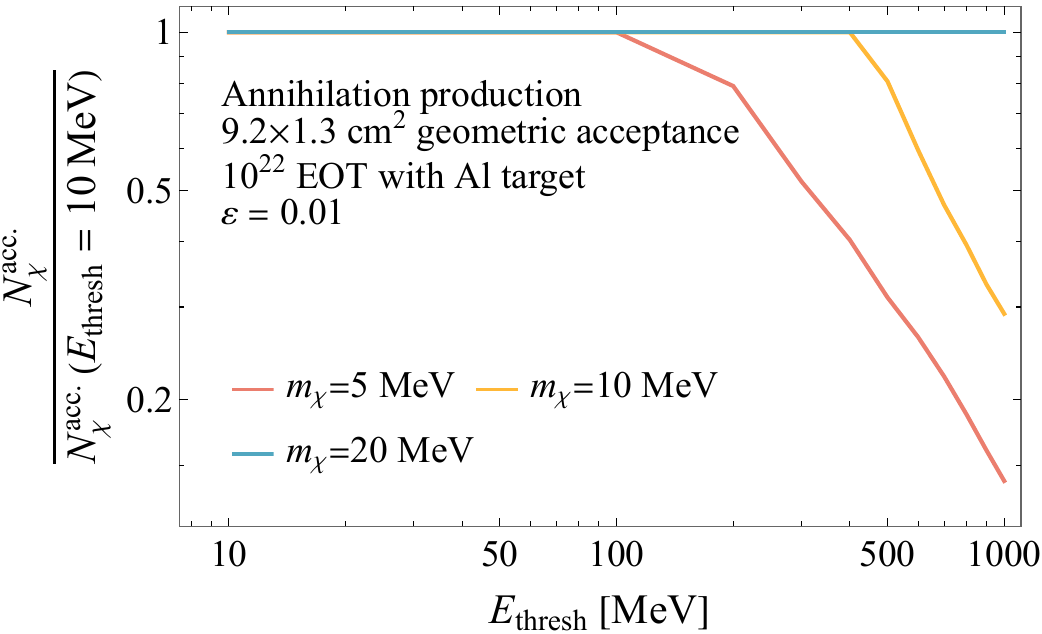}
    \caption{ 
     \textbf{Left:} The number of mCPs produced in electron-positron annihilation in the EM shower that pass through a detector of area $9.2 \times 1.3~{\rm cm}^2$ located 20~m downstream of the target ($N_\chi^\text{acc.}$) versus the energy threshold, for a millicharge of $\varepsilon=0.01$.  
    \textbf{Right:}
    The ratio of $N_\chi^\text{acc.}$ to The ratio of $N_\chi^\text{acc.}$ to    $N_\chi^\text{acc.}(E_{\rm thresh} = 10~\text{MeV})$, the accepted number of mCPs evaluated at a shower energy threshold of 10~MeV. We see that for $E_{\rm thresh}\lesssim 100~{\rm MeV}$, the flux hitting the detector is insensitive to the shower threshold. 
    \label{fig:annihilation_shower_number}}
\end{figure}

The annihilation process dominates the mCP flux arriving at the detector in the $2~\text{MeV}\lesssim m_\chi \lesssim 50~\text{MeV}$ window (see right panel of \cref{fig:shower_production_number}). 
In \cref{fig:annihilation_shower_number}, we show that an energy threshold of 10~MeV is sufficient to reliably estimate all the mCPs produced from secondary particles in the EM shower, which pass through the downstream detector area of mCP masses between 5~and 50~MeV. 
The kinematic threshold for producing mCPs with mass $m_\chi=5$~MeV is 100~MeV; hence, the results are independent of any threshold choice below 100~MeV. For much lower $m_\chi$, although the kinematic threshold is lower, the mCPs produced from low-energy positron will have a wider angular distribution and will not pass through the downstream detector.  Moreover, as shown in the right panel of  \cref{fig:shower_production_number}, the annihilation process is not competitive to the trident process in the low mass region, so that a 10~MeV threshold is sufficiently low also to estimate the mCPs produced in electron-positron annihilation.  Similar conclusions hold for the Compton production, which in any case is subdominant. 

\section{Multiple scattering \label{app:MCS} }
From the beam-dump to the well where the detector is located, there is 5.4~m of concrete and 14.2~m of dirt~\cite{Battaglieri:2020lds}. When mCP particles pass through this material, they can be deflected by multiple small-angle Coulomb scatters. The theory of multiple Coulomb scattering is well developed, and the core of the  distribution in scattering angle can be well modeled by a Gaussian distribution with a root-mean-squared (RMS) width given by~\cite{ParticleDataGroup:2024cfk},
\begin{equation}
    \theta_0 = \frac{13.6~\text{MeV}}{\beta p_\chi}\varepsilon\sqrt{\frac{x}{X_0}}\left(1+0.088\log_{10}\left(\frac{x\varepsilon^2}{X_0\beta^2}\right)\right)\,,
\end{equation}
where $\beta$ is the velocity, $p_\chi$ is the momentum of the mCP particle from the beam dump, $\varepsilon$ is its millicharge, $x$ is the thickness of the medium, and $X_0$ is the medium's radiation length.  The radiation length of concrete is 11.55~cm~\cite{concrete}. We model the dirt as standard rock (with radiation length of 26.54 ${\rm g}/{\rm cm}^2$~\cite{standardrock}), accounting for the decreased density  of dirt (1.7~$\text{g/cm}^3$ for dirt~\cite{BDX:2017jub} vs 2.7~$\text{g/cm}^3$ for standard rock), and we obtain $X_0 = 15.9~{\rm cm}$ for the dirt. We add the RMS deflecting angle from concrete and dirt in quadrature for the total result, since both materials are homogeneous and appear in series. 

\begin{figure}[t!]
    \centering
    \includegraphics[width=0.55\textwidth]{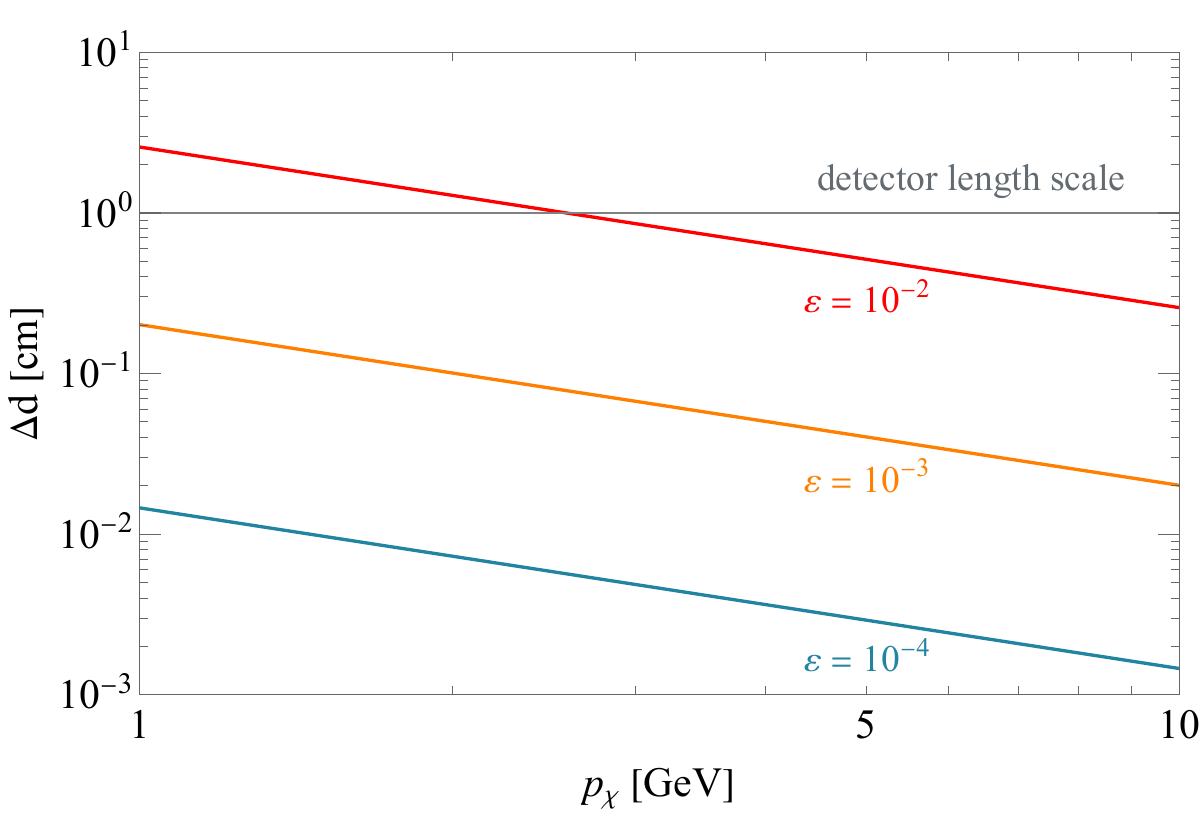}
    \caption{The deflected distance from multiple scattering on the plane of a detector located 20~m from the beam dump (after traversing 5.4~m of concrete and 14.2~m of dirt) as a function of mCP momentum, $p_\chi$, for different millicharges, $\varepsilon$. A minimum detector length scale of 1~cm is marked as the gray line.  We see that for the values of $\varepsilon$ that are not presently excluded, multiple Coulomb scattering does not affect the sensitivity projections computed in the main text.}
    \label{fig: multi_scattering}
\end{figure}

To study the impact on detection prospects with a Skipper-CCD, we compute the transverse distance of the  particles on the plane of detector (located $20~{\rm m}$ away) caused by scatterings in the concrete and dirt using $\Delta d = \theta_0\times20~\text{m}$. The deflection depends on the millicharge and momentum of the mCP (we assume $\beta = 1$, which is very accurate for $m_\chi\lesssim1$~GeV). We plot our results as a function of $p_{\chi}$ for different values of millicharge in ~\cref{fig: multi_scattering}. One immediately sees that the deflection by the dirt and concrete is not experimentally significant for $\varepsilon\lesssim10^{-2}$. This region of (relatively large) $\varepsilon$ is already robustly excluded by existing experiments as can be seen in~\cref{fig:pure_mCP_plot}. We therefore conclude that the multiple scatterings in the medium does not have a significant impact on our projected sensitivity, except for the regions of large $\varepsilon$ that are already excluded by existing searches.

\section{Beam-induced neutron backgrounds\label{app:backgrounds}}
The BDX collaboration studied beam-related and cosmic-ray backgrounds for the proposed BDX-DRIFT-1m and BDX-DRIFT-10m detectors (this is distinct from the proposed BDX PbWO$_4$ detector), concluding that a zero-background experiment is possible with an active veto system~\cite{SnowdenIfft2019BeamDumpDM, BDX:2016akw}. In particular, background particles such as neutrinos, muons, neutrons, and photons were simulated using GEANT4 with threshold energies of 10~MeV. However, a Skipper-CCD detector will (a) not have an active veto system and (b) has an energy threshold of $O({\rm eV})$ compared to the higher thresholds for the BDX-DRIFT design. Therefore, it is important to carefully re-consider background particles, which may have energies that are undetectable in the proposed BDX experiment and were thus not simulated in their studies. 

In general, particles with energy of $O({\rm MeV})$ will deposit large amount of charge in the Skipper-CCD detector (much more than the few electron-hole pairs from mCPs), which can be masked. Cosmic-ray background as well as beam-induced neutrinos, muons, and photons will all be highly energetic, and therefore these events can be easily rejected. As discussed in \cref{sec:backgrounds}, the required masking of high-energy backgrounds will decrease the effective exposure, which we include as an efficiency factor in our final sensitivity estimates. In the remainder of this appendix, we will discuss beam-induced neutrons that reach the detector, since these have been estimated by BDX. (Note that cosmic-ray-induced neutrons near the detector are likely a larger concern and require careful study by the proponents of a particular experiment, as they depend sensitively on the precise detector technology being used.)

There are two categories of beam-related neutrons that naively may contribute as a background: beam-induced neutrons produced from the electron interactions, and beam-induced muons that interact with the 10~m iron absorber (located along the beam line in between the concrete target area and the detector area) and that produces spallation neutrons. Neutrons as well as other muon secondaries are considered in~\cite{BDX:2016akw}, albeit with a discussion that assumes the BDX energy thresholds.

The recoil energy of a nucleus after an elastic neutron-nucleus scatter (assuming non-relativistic kinematics) is given by 
\begin{equation}
    \label{eq:TA-rec-neutron-scatter}
    T_{A,{\rm rec}} = \frac{4 \ T_n \ m_n \ m_A \ \sin^2\left(\frac{\theta}{2}\right)}{(m_n+m_A)^2}\, ,
\end{equation}
where $T_{A,{\rm rec}}$ is the kinetic energy of the final state nucleus, $m_n$ ($m_A$) is the neutron (nucleus) mass, $\theta$ is the scattering angle in the lab-frame, and $T_n$ is the incoming neutron kinetic energy. 
In order for a neutron to deposit an ionization energy that could fake a mCP event, it would need to produce 4-7 electron-hole pairs, which corresponds to a narrow range of $T_{A,{\rm rec}}$; the precise range is uncertain due to the uncertainty in the ionization yield of low-energy nuclear recoils, but it (very) roughly corresponds to 150 to 400~eV~\cite{SuperCDMS:2023geu,Essig:2018tss,Sarkis:2022pvc}. 
The relevant neutron energies that can produce such $T_{A,{\rm rec}}$ depends sensitively on $\theta$, but can be as low as $\sim$1~keV as can be checked explicitly using \cref{eq:TA-rec-neutron-scatter}. 

The BDX collaboration estimates that for $10^{22}$~EOT, about 290 neutrons scatter in their proposed BDX-DRIFT-10~m detector, which has a volume of $1~{\rm m} \times 1~{\rm m} \times10~{\rm m}$~\cite{SnowdenIfft2019BeamDumpDM}. The ultralow-threshold sensors would be significantly smaller in size, so that the number of neutrons that pass through the sensor is significantly smaller; for example, naively extrapolating the BDX-DRIFT-10~m detector area in line with the beam (1~m$^2$) to the area of a single Skipper-CCD that is in line with the beam ($\sim$12~cm$^2$) suggests that the number of neutrons that pass through the Skipper-CCD sensor is $<1$.  In addition, even if it scatters, the probability of producing $Q=3e^--7e^-$ is low. 

Finally, we consider spallation neutron backgrounds from beam-induced muon interactions in the  iron absorber. The BDX collaboration considered beam-induced muon secondaries and carefully simulates neutron production from muons three times with increased statistics~\cite{BDX:2016akw}. It is concluded that neutron secondaries are a negligible background if the detector is placed 20 meters from the beam dump. This conclusion holds for both the nominal BDX detector and the smaller ultralow-threshold detectors we have proposed in this work.

\bibliographystyle{utphys}
\bibliography{references}

\end{document}